\documentclass[a4paper,9pt]{article}

\usepackage{anysize}
\marginsize{50pt}{40pt}{0pt}{60pt}
\usepackage{placeins}
\usepackage[version=4]{mhchem}
\usepackage{siunitx}
\DeclareSIUnit\Molar{M}
\usepackage[utf8]{inputenc}
\usepackage{graphicx}
\usepackage{xcolor}
\usepackage{array}
\usepackage{multirow}
\usepackage{float}
\usepackage{amsmath}
\usepackage{amsfonts}
\usepackage{amssymb} 
\usepackage{bbm}
\usepackage{booktabs}
\usepackage{leftindex}
\usepackage{microtype}

\usepackage[printonlyused]{acronym}
\acrodef{DCM}{directed configuration model}
\acrodef{ER}{Erd\H{o}s-R{\'e}nyi}
\acrodef{SVD}{singular value decomposition}
\acrodef{SAS}{\textit{singular angle similarity}}
\acrodef{GT}{ground truth}

\definecolor{CC_blue}{RGB}{16, 38, 148}  
\usepackage[colorlinks=true,
            urlcolor=CC_blue,
            citecolor=CC_blue,
            linkcolor=CC_blue,
            bookmarks=true,
            breaklinks]{hyperref}

\usepackage[                
        backend=bibtex,      
        style=numeric,      
        natbib=true,        
        hyperref=true,      
        alldates=year,      
	    uniquelist=false,   
        sorting=none,
	    isbn=false,
	    eprint=false,
	    doi=false,
        url=false,
	    maxcitenames=2,
        giveninits=true
]{biblatex}

\renewbibmacro{in:}{%
  \ifentrytype{article}{}{%
    \printtext{\bibstring{in}\intitlepunct}%
  }
}

\AtEveryBibitem{
    \clearfield{urlyear}
    \clearfield{urlmonth}
    \clearlist{language}
}

\begin{document}

\twocolumn[{\centering{\Huge Assessing the similarity of real matrices with arbitrary shape \par}\vspace{3ex}
	{\Large 
    Jasper Albers\textsuperscript{1,2*\dag},
    Anno C. Kurth\textsuperscript{1,2*\dag},
    Robin Gutzen\textsuperscript{1},
    Aitor Morales-Gregorio\textsuperscript{1,3}, 
    Michael Denker\textsuperscript{1}, 
    Sonja Gr\"{u}n\textsuperscript{1,4,5}, 
    Sacha J. van Albada\textsuperscript{1,6}, 
    Markus Diesmann\textsuperscript{1,4,7,8}\par}\vspace{2ex}
    \small 
    \textsuperscript{1}Institute for Advanced Simulation (IAS-6), J\"{u}lich Research Centre, J\"{u}lich, Germany.
    \textsuperscript{2}RWTH Aachen University, Aachen, Germany.
    \textsuperscript{3}Faculty of Mathematics and Physics, Charles University, Prague
    \textsuperscript{4}JARA-Institut Brain Structure-Function Relationships (INM-10), J\"{u}lich Research Centre, J\"{u}lich, Germany.
    \textsuperscript{5}Theoretical Systems Neurobiology, RWTH Aachen University, Aachen, Germany.
    \textsuperscript{6}Institute of Zoology, University of Cologne, Cologne, Germany.
    \textsuperscript{7}Department of Psychiatry, Psychotherapy and Psychosomatics, School of Medicine, RWTH Aachen University, Aachen, Germany.
    \textsuperscript{8}Department of Physics, Faculty 1, RWTH Aachen University, Aachen, Germany.\\ \vspace{2ex} 
    \textsuperscript{*}These authors contributed equally to this work.\\
    \textsuperscript{\dag} \href{mailto:j.albers@fz-juelich.de}{j.albers@fz-juelich.de}, \href{mailto:a.kurth@fz-juelich.de}{a.kurth@fz-juelich.de}\\
    \vspace{2ex}}
Assessing the similarity of matrices is valuable for analyzing the extent to which data sets exhibit common features in tasks such as data clustering, dimensionality reduction, pattern recognition, group comparison, and graph analysis. Methods proposed for comparing vectors, such as cosine similarity, can be readily generalized to matrices. However, this approach usually neglects the inherent two-dimensional structure of matrices. Here, we propose \ac{SAS}, a measure for evaluating the structural similarity between two arbitrary, real matrices of the same shape based on singular value decomposition. After introducing the measure, we compare \ac{SAS} with standard measures for matrix comparison and show that only \ac{SAS} captures the two-dimensional structure of matrices. Further, we characterize the behavior of \ac{SAS} in the presence of noise, as a function of matrix dimensionality, and when singular values are degenerate. Finally, we apply \ac{SAS} to two use cases: square non-symmetric matrices of probabilistic network connectivity, and non-square matrices representing neural brain activity. For synthetic data of network connectivity, \ac{SAS} matches intuitive expectations and allows for a robust assessment of similarities and differences. For experimental data of brain activity, \ac{SAS} captures differences in the structure of high-dimensional responses to different stimuli. We conclude that \ac{SAS} is a suitable measure for quantifying the shared structure of matrices with arbitrary shape.
\medbreak
\par\vspace{5ex}]


\section{Introduction}
Social groups, transportation systems, chemical reactions, brains - many complex systems are governed and commonly characterized by the pairwise interactions of their constituent elements. Typically, these interactions are described by matrices that can represent covariance structures, spatio-temporal dependencies, or the connections and interactions in a network, forming the foundation for the mathematical treatment of such complex systems \citep{Newman03_167}. Common examples include genetic variance \citep{Calsbeek09_2627}, ecological food chains \citep{Sol2001_2039}, or stock markets \citep{Piccardi2011_35}. Additionally, many other types of structured data can be represented in matrix form, ranging from test scores for groups of subjects to parallel time series data. Quantifying the similarity between such matrices is important for distinguishing features of the underlying systems. Examples include the representations of stimuli in artificial and biological neural networks. The extent to which such representations are similar helps to address questions regarding convergent learning, i.e., whether solutions found by neural networks are universal or specific. Representational similarity has been investigated across different neural network architectures \citep{Morcos18, Kornblith19_3519}, different training data \citep{Kornblith19_3519}, and different initializations of the parameters \citep{Li15_196}. In the context of biological neural networks, differences in the representation within brain areas \citep{Haxby01_2425}, between species \citep{Kriegeskorte08_1126} as well as between brains and artificial neural networks \citep{Yamins13, Sussillo15_1025, Du24} have been studied. See \citep{Sucholutsky23} for a review of the topic.

Classical measures of the similarity between two matrices $A$ and $B$ are often based on the Frobenius scalar product $\langle A, B\rangle_{F} = \mathrm{tr}(AB^{T})$, leading to the Frobenius norm $||A -B||^{2}_{F}~=~\langle A-B,A-B\rangle_{F}$, or the cosine similarity $\langle A, B \rangle_{F}$ where $||A||_{F}=||B||_{F}=1$ \citep{Robert76_257}. However, the Frobenius norm and cosine similarity only take into account the numerical values of corresponding entries of the matrices and ignore their two-dimensional structure. For non-negative matrices, information theory-inspired approaches have been suggested \citep{Cichocki07_142, Cichocki10_1532}. These are conceptually similar in the sense that also here the relative position of matrix entries is ignored.

Measures partially overcoming the defect of ignoring the two-dimensional structure of matrices include the centered kernel alignment \citep{Cortes12_795, Kornblith19_3519} (CKA) and the normalized Bures similarity \citep{Tang20}. For linear kernels, CKA reads as $||AB^{T}||^{2}_{F} / ||AA^{T}||_{F}||BB^{T}||_{F}$. Other popular choices for CKA kernels include radial basis functions. The normalized Bures similarity is identical to CKA for linear kernels with the exception that the Frobenius norm is replaced by the nuclear norm. Kernel-based approaches \citep{Gartner03_129, Borgwardt05_8, Shervashidze11, Gretton12_723} are also used for assessing the similarity between graphs (represented by their adjacency matrices, see \citet{Kriege20_1} for a recent survey).

Other authors propose comparisons based on correlation coefficients or the coefficient of determination as a means to assess similarity \citep{Haxby01_2425, Kriegeskorte08_1126}. This approach is not limited to standard Pearson correlation coefficients but instead can also employ regularized regression analyses such as Ridge or LASSO regression.

More involved approaches reformulate ideas borrowed from canonical correlation analysis (CCA) \cite{Hotelling36_321} in the context of the comparison of two matrices \cite{Golub95_27}. They rely on the canonical angles between the subspaces spanned by the columns of the matrices. Here, problems may arise if the embedding space (the number of rows) is of similar size as the number of columns and, at the same time, the subspaces spanned by these columns are high-dimensional (relative to the dimension of the embedding space). In this case, the canonical angles cannot meaningfully distinguish the subspaces, limiting the applicability of these approaches. Alternatively, similarity measures based on Procrustes analysis have been proposed \citep{Andreella23_867}. Assuming that matrices $A$ and $B$ are centered, the Procrustes problem asks for the minimum of $||OA - B||_{F}$ where $O$ is an orthogonal matrix \citep{Schonemann66}. This minimum can be interpreted as a similarity and has a direct geometric meaning when viewing the matrix elements as coordinates of points in space: the Procrustes similarity assesses to which extent two objects, represented by the matrices $A$ and $B$, overlap after reflecting and rotating them so that the overlap is maximized. Other formulations consider a measure like the cosine similarity after having applied the optimal orthogonal transformation \citep{Williams21_4738}. See \citet{Ding21_1556} or \citet{Cloos24} for a comparison of some of the mentioned similarity measures. Finally, for symmetric positive-definite matrices (SPD, e.g. covariance matrices), similarity measures based on geodesic distances induced by a natural Riemannian metric on the space of SPD matrices have been studied \citep{Barachant11_920, Ju22_10855}. The induced distance is invariant under the action of the general linear group on the SPD matrices and thus especially under joint diagonalization.

Previous work addresses the comparison of symmetric matrices using eigenangles\textemdash the angles between eigenvectors of the compared matrices \citep{Gutzen23_104813}. Small eigenangles indicate a good alignment of the respective eigenspaces, enabling the definition of a similarity score. The authors employ an analytical description of the similarity score based on random matrix theory to devise a statistical test for the comparison of such matrices.
Asymmetric matrices usually have complex eigenvectors and eigenvalues, making their ordering ambiguous. Thus, an extension of the eigenangle test to asymmetric matrices is not straightforward. Additionally, the approach implicitly assumes that all eigenvalues are non-degenerate. Finally, the eigenangle test is by definition not applicable to non-square matrices. In this study, we overcome these limitations by proposing a refined matrix similarity measure that naturally extends to the comparison of any two real matrices with identical shapes. Using singular value decomposition (SVD) instead of eigendecomposition, we derive \ac{SAS} from the respective left and right singular vectors and singular values.
In this way, we vastly generalize the approach introduced in \citep{Gutzen23_104813} and enable a multitude of applications not possible previously.

In \autoref{sec:methods}, we formally define \ac{SAS} and derive basic properties. Further, three types of matrices are introduced that we use in the following for the evaluation of SAS and comparison to other similarity measures: random matrices with continuously distributed entries (\autoref{sec:random_matrices}), adjacency matrices of random graphs (\autoref{sec:random_graphs}), and massively parallel neural activity recordings (\autoref{sec:brain_data}). Finally, we detail how degeneracy is gradually introduced to the singular value spectrum of matrices (\autoref{sec:degenerate_matrices}).

We start our analysis by providing a geometric interpretation of \ac{SAS} (\autoref{sec:geometric_interpretation}). Next, we compare \ac{SAS} with standard similarity measures for matrices: on the basis of generic random matrices we show that only \ac{SAS} captures certain salient two-dimensional correlation structures (\autoref{sec:comparison_other_metrics}). This comparison follows a templated structure where we first calculate the self-similarity distribution of different samples of one random matrix class assessed by \ac{SAS}. This distribution is then compared with the cross-similarity distribution of \ac{SAS} scores from sampled random matrices of the given class with other random matrix classes. Third, we characterize the behavior of \ac{SAS} under changes in dimensionality, perturbation of entries, and degeneracy of singular values (\autoref{sec:characterization}). Fourth, we evaluate the similarity across instances of six probabilistic graph models that are commonly used to describe network architecture (\autoref{sec:categorization_random_graphs}). With this use case, we demonstrate that \ac{SAS} is able to differentiate between the connectivity in network graphs by means of their adjacency matrices. Finally, we apply \ac{SAS} on experimental data, evaluating the similarity of brain activity in the visual cortex of macaques in response to four different visual stimuli (\autoref{sec:separation_brain_states}). This application shows that \ac{SAS} can identify underlying features in the presence of realistic noise. 

In conclusion (\autoref{sec:discussion}), we show that \ac{SAS} is a well-behaved measure for structural similarity in matrices that is applicable in different scientific domains. It highlights shared variability between matrices and allows for a distinction of models or processes underlying their generation.

\section{Methods}\label{sec:methods}
\subsection{Singular angle similarity}\label{sec:similarity}
\begin{figure}[ht]
    \begin{center}
        \includegraphics[width=\linewidth]{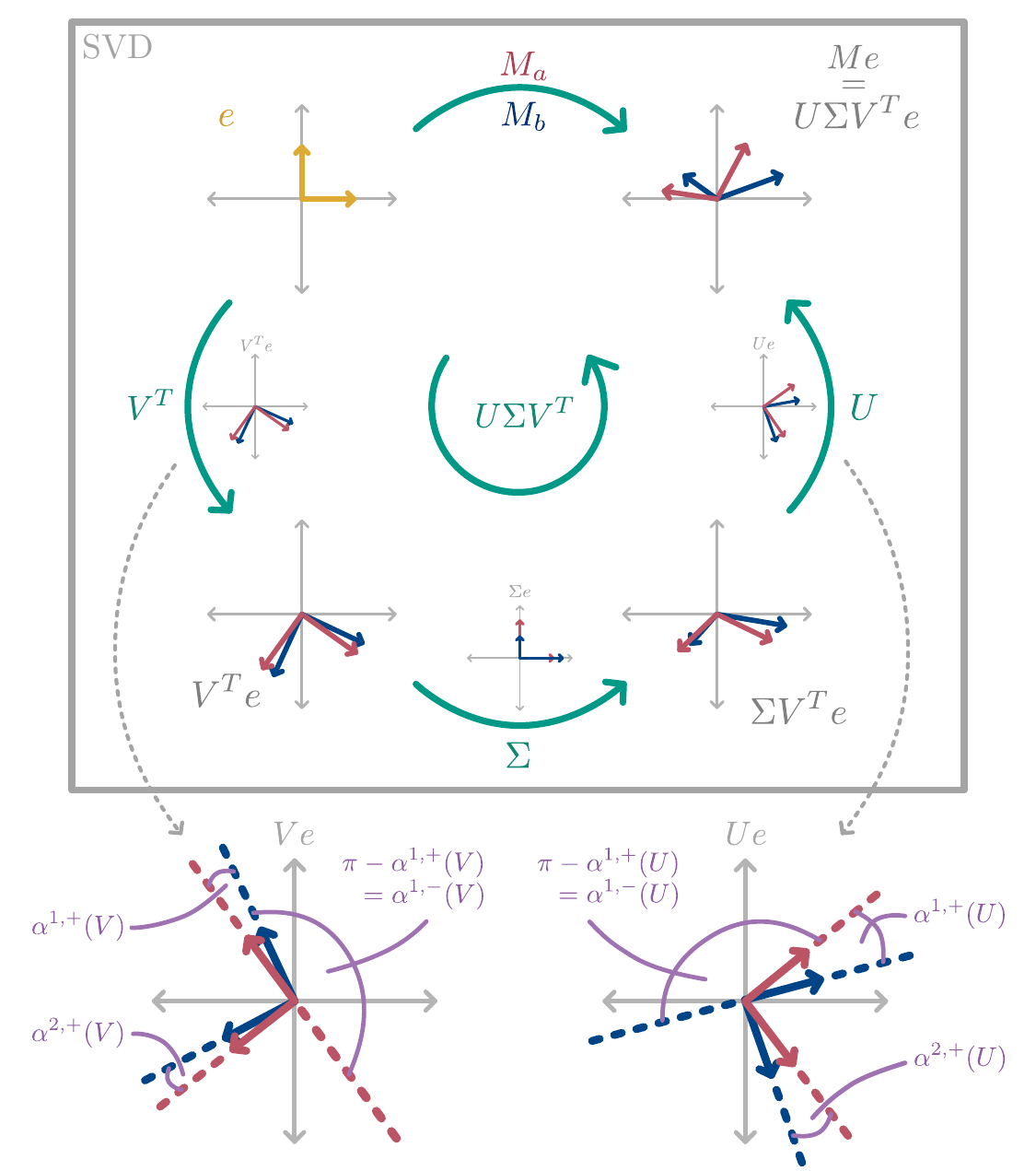}
    \end{center}
    \caption{
    \textbf{Singular value decomposition and singular angles.}
    Schematic representation of the transformations applied on the basis vectors $e$ (yellow) by the components of \ac{SVD} for two $2 \times 2$ matrices $M_a$ (red) and $M_b$ (blue). Small graphs next to the green arrows illustrate the isolated action of the corresponding transformations $V^T$, $\Sigma$, and $U$ on $e$.  Below, the singular angles of the first ($\alpha^{j, +}(V)$ and $\alpha^{j, +}(U)$) and second ($\alpha^{j, -}(V)$ and $\alpha^{j, -}(U)$) kinds are shown as the angles between the column vectors of $V_a$ and $V_b$, and $U_a$ and $U_b$, respectively, as defined in \autoref{eq:singular_angle_first_kind}.}
    \label{fig:erklaerbaer}
\end{figure}

To assess the similarity of two arbitrary, real, $m \times n$ matrices $M_{a}, M_{b}$, we devise a measure based on \ac{SVD} \citep{Trefethen_NLA}.
Without loss of generality, we assume $m \leq n$.
\ac{SVD} guarantees the existence of orthogonal matrices $U_{i}\in \mathbb{R}^{m\times m}$ and $V_{i} \in \mathbb{R}^{n\times n}$, and diagonal matrices $\Sigma_{i}=\mathrm{diag}(\sigma_{i}^{1}, ..., \sigma_{i}^{m}) \in \mathbb{R}^{m \times n}$ where $\sigma_{i}^{j} \geq \sigma_{i}^{l}\geq 0 $ for $l > j \geq 1$ such that
\begin{equation}
    M_{i} = U_{i} \Sigma_{i} V_{i}^{T}~.
    \label{eq:SVD_eq}
\end{equation}
Here, $i\in\{a, b\}$.
\ac{SVD} is schematically presented for a $2 \times 2$ case in \autoref{fig:erklaerbaer}.
The columns of $V_{i}$, denoted by $v_{i}^{1}, ..., v_{i}^{n}$, are the right singular vectors, and the columns of $U_{i}$, denoted by $u_{i}^{1},...,u_{i}^{m}$, are the left singular vectors.
With this, the \ac{SVD} can also be written in the form
\begin{equation}
    M_{i} = \sum_{j=1}^{m} \sigma_{i}^{j} u_{i}^{j} \otimes v_{i}^{j}~.
\end{equation}
Here, $\otimes$ denotes the outer product of two vectors.
Thus, under the action of $M_{i}$ the vectors $v^{j}_{i}$ are transformed into the vectors $\sigma_{i}^{j}u_{i}^{j}$.
The singular values $\sigma_{i}^{1}, ..., \sigma_{i}^{m}$ are unique\textemdash they are the square root of the eigenvalues of $M_{i}M_{i}^{T} \in \mathbb{R}^{m \times m}$.
We note that if both $M_{i}$ are symmetric positive-definite matrices (e.g., covariance matrices), then $U_{i}=V_{i}$, and the singular values become the eigenvalues.

For simplicity of the derivations, we at first assume that there are no degenerate singular values except zero.
The overwhelming majority of higher-dimensional matrices encountered in practice indeed satisfy this assumption.

Left and right singular vectors that correspond to non-degenerate singular values are uniquely determined up to a joint multiplication by $-1$. Thus, the vector pairs $(u_{i}^{j}, v_{i}^{j})$ and $(-u_{i}^{j}, -v_{i}^{j})$ both are equally valid singular vectors to a non-degenerate singular value $\sigma_{i}^{j}\geq 0$.

If $\min \{\sigma_{a}^{j}, \sigma_{b}^{j}\} \neq 0$ we define the left singular angle $\alpha^{j, +}(U)$ and the right singular angle $\alpha^{j, +}(V)$ of the first kind as 
\begin{align}
    \alpha^{j, +}(U) & = \sphericalangle(u_{a}^{j}, u_{b}^{j}) = \arccos\left( \langle u_{a}^{j}, u_{b}^{j} \rangle\right), \notag\\
    \quad
    \alpha^{j, +}(V) & = \sphericalangle(v_{a}^{j}, v_{b}^{j}) = \arccos\left( \langle v_{a}^{j}, v_{b}^{j} \rangle\right). 
\label{eq:singular_angle_first_kind}
\end{align}
Due to the ambiguity in vector pairs of left and right singular vectors we additionally define left singular angles of the second kind as $\alpha^{j, -}(U)=\sphericalangle(-u_{a}^{j}, u_{b}^{j})= \sphericalangle(u_{a}^{j}, -u_{b}^{j})$ and \textit{mutatis mutandis} for right singular angles of the second kind $\alpha^{j, -}(V)$.
The singular angles of the first and second kinds are visualized in \autoref{fig:erklaerbaer}.
There, we have
\begin{equation}
    \alpha^{j, +}(U)+ \alpha^{j, -}(U) = \pi =  \alpha^{j, +}(V)  + \alpha^{j, -}(V)~.
\end{equation}
Due to the ambiguity in the sign, one has to consider either $\left(\alpha^{j, +}(U), \alpha^{j, +}(V)\right)$ or $\left(\alpha^{j, -}(U), \alpha^{j, -}(V)\right)$ together. We define the \textit{singular angle} as the smaller average of the two choices
\begin{align}
    \alpha^{j} &= \min \left\{ \frac{\alpha^{j, +}(U) + \alpha^{j, +}(V)}{2}, \frac{\alpha^{j, -}(U) + \alpha^{j, -}(V)}{2} \right\} \notag \\
    &= \min \{ \bar{\alpha}^{j}, \pi - \bar{\alpha}^{j} \}
\end{align}
where $\bar{\alpha}^{j} = (\alpha^{j, +}(U) + \alpha^{j, +}(V)) / 2$.
Using the angular similarity
\begin{equation}
    \Delta^{j} = 1 - \frac{\alpha^{j}}{\pi/2} \in [0, 1]
    \label{eq:angle_similarity}
\end{equation}
and defining the \textit{singular value score} as $w^{j} = w(\sigma_{a}^{j}, \sigma_{b}^{j})$ where $w(x,y) \geq 0$ denotes a weight function, we calculate \ac{SAS} as the weighted average of the angular similarities: 
\begin{equation}
    \mathrm{SAS} = \frac{\sum_{j}^{k} w^{j} \Delta^{j}}{\sum_{j}^{k} w^{j}} \in [0,1]~.
    \label{eq:SAS}
\end{equation}
Here, $k$ is the largest natural number less than or equal to $m$ such that $\min \{\sigma^{k}_{a}, \sigma^{k}_{b}\} \neq 0$.
In the following, we choose
\begin{equation}
    w(x, y) = (x + y) / 2~.
\label{eq:weight_function}
\end{equation}
Other possible choices include $w(x, y)~=~\sqrt{(x^{2}~+~y^{2})/2}$ (c.f. \cite{Gutzen23_104813}) and $w(x,y)=\sqrt{x\cdot y}$.
One can substitute other vector-based similarity measures for the angular similarity defined in \autoref{eq:angle_similarity}.
For example, substituting cosine similarity yields $\Delta^j~=~\cos(\alpha^j)$.

According to our definition of \ac{SAS} in \autoref{eq:SAS}, singular angles stemming from singular vectors of which at least one has a corresponding singular value of zero do not contribute to \ac{SAS}.
By definition, \ac{SAS} can only be computed for matrices of the same shape.

\paragraph{Degenerate case}
If two or more nonzero singular values are identical, the described approach cannot be readily applied since there is no canonical choice for pairing singular vectors: left and right singular vectors $u_{i}^{k},...,u_{i}^{k+l}$ and $v_{i}^{k},...,v_{i}^{k+l}$ corresponding to the degenerate singular value $\sigma = \sigma_{i}^{k}=...=\sigma_{i}^{k+l}$ are only unique up to an orthogonal transformation acting on the subspaces spanned by the singular vectors. This is the higher-dimensional generalization of the ambiguity regarding the factor $-1$ for the one-dimensional case described above. Writing the corresponding vectors as columns of matrices $U_{i}^{[k:k+l]} \in \mathbb{R}^{n\times l}$ and $V_{i}^{[k:k+l]} \in \mathbb {R}^{m\times l}$, this means that the columns of the matrices
\begin{equation}
    U_{i}^{[k:k+l]} O \quad \textnormal{and} \quad V_{i}^{[k:k+l]} O
\end{equation}
are valid left and right singular vectors. Here, $O \in \mathbb{R}^{l \times l}$ is an arbitrary orthogonal matrix. Note that in order to maintain consistency between the singular vectors, the same orthogonal transformation $O$ must be applied to both, $U_{i}^{[k:k+l]}$ and $V_{i}^{[k:k+l]}$. Thus, left and right singular vectors can only be given up to this ambiguity, and consequently singular angles according to \autoref{eq:singular_angle_first_kind} are not well-defined.

This problem can be mitigated by using the \textit{canonical angles} between the subspaces spanned by the left and right singular vectors of the matrices $M_{i}$ \citep{Jordan75_103}.
If the two subspaces are given in terms of orthonormal bases $B_{i}$ (written again as columns of matrices), the canonical angles are the angles between corresponding column vectors of $B_{a} O_{a}$ and $B_{b} O_{b}$, where $O_{a}, O_{b}$ are suitable orthogonal transformations \citep{Zhu13_325} (see \autoref{supp:canonical_angles}). This can be interpreted as optimally aligning the two orthonormal coordinate systems while keeping the subspaces invariant.

We define singular angles for degenerate singular values building on that interpretation: assuming without loss of generality $\sigma = \sigma_{a}^{k}=...=\sigma_{a}^{k+l}$, we define $U$-aligned left and right singular angles for $k \leq j \leq k+l$ as 
\begin{align}
    &\alpha^{j, U}(U) = \sphericalangle\left(U_{a}^{[k:k+l]} O_{a}^{U}e_{j}, U_{b}^{[k:k+l]} O_{b}^{U}e_{j}\right) \notag \\
    &\alpha^{j, U}(V)= \sphericalangle\left(V_{a}^{[k:k+l]} O_{a}^{U}e_{j}, V_{b}^{[k:k+l]} O_{b}^{U}e_{j}\right)~.
\end{align}
Here, $e_{j}$ is the $j$-th standard normal basis vector, and the orthogonal transformations $O_{a}^{U}, O_{b}^{U}$ are chosen such that $\alpha^{j, U}(U)$ are the canonical angles. The $U$-aligned singular angles are then given by the mean of the $U$-aligned left and right singular angles:
\begin{equation}
    \alpha^{j, U} = \frac{\alpha^{j, U}(U) + \alpha^{j, U}(V)}{2}
\end{equation}
The corresponding angles in the $V$-aligned case are defined \textit{mutatis mutandis}. The singular angles are then given by either the $U$- or $V$-aligned singular angles, depending on which have the smaller sum:
\begin{equation}
    \alpha^j = \alpha^{j, X} \textnormal{ where } X = 
    \begin{cases}
        U &\textnormal{if } \sum_j \alpha^{j, U} < \sum_j \alpha^{j, V} \\
        V &\textnormal{if } \sum_j \alpha^{j, V} \leq \sum_j \alpha^{j, U}
    \end{cases}
\end{equation}
In this way, the singular angles for singular vectors corresponding to degenerate singular values directly generalize the singular angles in the non-degenerate case.
For their contribution to \ac{SAS}, we define $w^{j}=w\left(\sigma_{a}^{j}, \sigma_{b}^{j}\right)$ for $k \leq j \leq k+l$, i.e. the singular values are combined in their matching order.
A complication may arise when $\sigma_{a}^{k}=...=\sigma_{a}^{k+l}$ and $\sigma_{b}^{k-m}=...=\sigma_{b}^{k+n}$ for some $m$ and $n$. Here, the degenerate subspaces of the two matrices are partially overlapping. We treat this case by applying the above described method to the subspaces given by $U_{i}^{[k-m:k+\max\{n,l\}]}$, $V_{i}^{[k-m:k+\max\{n,l\}]}$.

Additionally, if two singular values are close so that small perturbations can lead to a change in their order, the pairing of singular vectors will change, potentially leading to different \ac{SAS} values. This can be avoided by rounding the singular values to a precision determined for the matrices at hand. Thereby, degeneracy is introduced which can be treated as described above.

\paragraph{Large matrices}
For applications, it might be impractical to compute the full SVD of the matrices $M_{a}, M_{b}$ if $n,m$ are large. In this case, \ac{SAS} can still be used when replacing the full SVD with the truncated SVD \citep{Halko11_217}: instead of decomposing the matrices exactly according to \autoref{eq:SVD_eq}, one seeks to find a low-rank approximation:
\begin{equation}
    M_{i} \approx \leftindex^{t}U_{i} \leftindex^{t} \Sigma_{i} \leftindex^{t} V_{i}^{T}
    \label{eq:tSVD_eq}
\end{equation}
Here, $\leftindex^{t} U_{i} \in \mathbb{R}^{m \times k}$ and $\leftindex^{t} V_{i} \in \mathbb{R}^{n \times k}$ have orthogonal columns and $\leftindex^{t} \Sigma_{i} \in \mathbb{R}^{k \times k}$ is a diagonal matrix with non-negative entries. The singular values and corresponding singular vectors of the truncated SVD are identical to the $k$ largest singular values of the full SVD and their corresponding singular vectors. The order of the approximation, $k$, depends on the concrete application. Note that the Eckart-Young-Mirsky theorem asserts that this low-rank approximation is minimal with respect to the Frobenius norm given the order $k$ \citep{Golub87_317}. The decomposition in \autoref{eq:tSVD_eq} can straightforwardly be used to compute \ac{SAS}. The resulting similarity between matrices is still informative if the singular values decrease in magnitude sufficiently quickly. The validity of this assumption depends on the matrices at hand. In practice, oftentimes only few singular values, compared to the dimension of the matrices, are of large magnitude.

\paragraph{Matrices of different shapes}
\label{sec:diff_shapes}
By definition, \ac{SAS} can only be applied to matrices of identical shape. This poses limitations for situations in which matrices of different shapes occur naturally. Examples include data matrices from neuroscientific experiments involving behavioral paradigms where the time needed by a test animal for the completion of a task varies between trials and subjects. While such situations can in principle be addressed with various forms of ``time-warping" (see e.g. \citep{Williams20_246} for the domain of neuroscience), similar transformations leading to satisfactory results may not exist in all contexts. To nonetheless assess the similarity, \ac{SAS} can be extended by zero-padding the singular vectors with smaller dimensionality so that the scalar products in \autoref{eq:singular_angle_first_kind} remain well-defined. Thereby, dimensions only present in the vectors with larger dimensionality are neglected.

\subsection{Random matrices}\label{sec:random_matrices}
To compare \ac{SAS} to standard measures of matrix similarity, we define the following classes of random matrices with shape $N \times N$ where each entry is drawn from a continuous probability distribution. Such matrix ensemble are of great theoretical interest and of widespread use in various domains of science \citep{PottersBouchaud20, Tao12}.  Numerical values for the corresponding model parameters are summarized in \autoref{tab:parameters_matrices}.

\paragraph{Uncorrelated normal matrix (UC)}\label{paragraph:uncorr_normal}
For random matrices of this class, each entry is drawn independently from a normal distribution with the same mean $\mu$ and variance $\sigma^{2}$. Collections of these matrices are also referred to as real Ginibre Ensemble \citep{Ginibre65_440}.

\paragraph{Cross-correlated normal matrix (CC)}
We first independently sample $N$ random vectors from an $N$-dimensional normal distribution with mean $\mu$ and covariance matrix $C$ where
\begin{equation*}
    C_{ij} = \frac{a}{N} \cdot \exp\left(-b\frac{|i-j|}{N}\right)~.
\end{equation*}
Thus, the entries of the correlation matrix decay exponentially with distance from the diagonal.
The sampled vectors are the columns of an $N\times N$ matrix $K^{1}$. We repeat the process to obtain an independent matrix $K^{2}$ to finally define $K = (K^{1} + {K^{2}}^{T}) /2$.
Since each entry of the $K^{i}$ is normally distributed, so is each entry of their sum $K$, and the covariance between entries $K_{kl}$ and $K_{mn}$ is $(C_{kl} + C_{mn}) / 4$. Normalization by $N$ in the argument of the exponential ensures that the strength of the correlation scales with the size of the matrix.

\paragraph{Cross-correlated block matrix (CB)}
Again, uncorrelated normal (UC) matrices $B$ are sampled. Then, the entries $B_{kl}$ where $b^{\mathrm{lower}} \leq k \leq b^{\mathrm{upper}}$ and $b^{\mathrm{lower}} \leq l \leq b^{\mathrm{upper}}$ are replaced by a correlated normal structure as defined above, forming a block on the diagonal.

\paragraph{Shuffled, cross-correlated block matrix (SB)}
Matrices are sampled according to CB. Then, the rows are permuted randomly while the columns remain untouched.

\paragraph{Doubly shuffled, cross-correlated block matrix (DB)}
Matrices are sampled according to CB. Then, the rows and columns are permuted randomly.

\begin{table}[ht]
\footnotesize
\begin{tabular}{@{}llll@{}}
\toprule
Parameter            & Model(s) & Meaning                               & Value \\ \midrule
$N$             & all      & dimensionality                       & $300$   \\
$\mu$             & all      & mean of distribution& $0$ \\
$\sigma^{2}$               & UC       & variance of distribution   & $1/N$    \\
$a$               & CC       & peak covariance & $10$    \\
$b$               & CC       & inverse characteristic length & $100$    \\
$b^{\mathrm{lower}}$             & CB       & block index & $10$    \\
$b^{\mathrm{upper}}$             & CB       & block index & $90$    \\\bottomrule 
\end{tabular}
\centering
\caption{Parameters of random matrices: uncorrelated normal matrix (UC), cross-correlated normal matrix (CC), cross-correlated block matrix (CB), and shuffled, cross-correlated block matrix (SB).}
\label{tab:parameters_matrices}
\end{table}

\subsection{Random graphs}
\label{sec:random_graphs}
We compare the adjacency matrices of six well-known network models with \ac{SAS}.
For all graphs, we derive the parameters such that the mean total number of connections $N_c$ in the graph is conserved.
\autoref{tab:parameters} summarizes the numerical values chosen for the corresponding model parameters.

\paragraph{Erd\H{o}s-R{\'e}nyi (ER)}
In this network model \cite{Erdos59}, every connection has the same probability of being realized: $p~=~\frac{N_c}{N_e}$, where $N_e$ is the total number of possible connections in the graph. Note that this network model maximizes the entropy under the constraint that the mean number of connections is fixed \citep{Park04_066117}.

\paragraph{Directed configuration model (DCM)}
In a directed configuration model \citep{Cooper04_319}, a two-step probabilistic process is applied. First, random indegrees and outdegrees are drawn for each node such that the total number of connections across nodes is preserved (we fix these numbers for all graph instances). Second, connections are established by randomly matching each outgoing connection with an incoming connection. Thus, two nodes can have more than one connection, and the resulting adjacency matrix is not strictly binary.

\paragraph{One-cluster Erd\H{o}s-R{\'e}nyi (OC)}
Based on an \ac{ER} graph, we introduce a single cluster by increasing $p$ between a certain subset of nodes of the network, while uniformly decreasing $p$ for all other connections such that $N_c$ is conserved. The relative increase of $p$ is denoted by $r$, and the location of the cluster on the diagonal is defined by the bounding indices $b^{\mathrm{lower}}$ and $b^{\mathrm{upper}}$.

\paragraph{Two-cluster Erd\H{o}s-R{\'e}nyi (TC)}
For the two-cluster \ac{ER} network, we create two non-overlapping clusters on the diagonal using the same method as in the OC model. The nodes that form the clusters are chosen such that there is maximal overlap with the single cluster of the OC model: the first cluster starts at the same index $b^{\mathrm{lower}}$ and extends up to index $b^{\mathrm{mid}}$, and the second cluster starts at index $b^{\mathrm{mid}}+1$ and extends up to index $b^{\mathrm{upper}}$.

\paragraph{Watts-Strogatz (WS)}
We create a small-world network following \cite{Watts98}. Here, $N_n$ nodes initially form a ring, where each node is connected to $k=\frac{N_c}{N_e} (N_n - 1)$ of its nearest neighbors. Afterwards, all connections are uniformly redistributed with probability $p_{WS}$. Note that this model is undirected.

\paragraph{Barab\'{a}si-Albert (BA)}
As an example of a scale-free network, we create Barab\'{a}si-Albert networks as introduced in \cite{Albert02}.
Here, from an initial star graph with $m=\frac{N_c}{N_e} (N_n - 1) / 2$ nodes, new nodes are added subsequently until the desired number of nodes $N_n$ is reached. Each added node is connected to $m$ existing nodes, where the probability of each existing node being selected for a new connection is proportional to the number of connections it already has. Note that this model is undirected.\\

\begin{table}[ht]
\footnotesize
\begin{tabular}{@{}llll@{}}
\toprule
Param.            & Model(s) & Meaning                               & Value \\ \midrule
$N_n$             & all      & number of nodes                       & $300$   \\
$N_e$             & all      & number of possible connections        & $90000$ \\
$N_c$             & all      & mean number of connections            & $9000$  \\
$r$               & OC       & relative increase of $p$ in cluster   & $10$    \\
$b^{\mathrm{lower}}$  & OC & cluster index & $50$    \\
$b^{\mathrm{upper}}$  & OC & cluster index & $100$    \\
$b^{\mathrm{mid}}$     & TC & index between clusters                     & $90$    \\
$p_{\mathrm{WS}}$ & WS       & reconnection probability              & $0.3$   \\\bottomrule 
\end{tabular}
\centering
\caption{Parameters of random graphs: Erd\H{o}s-R{\'e}nyi (ER), directed configuration model (DCM), one cluster (OC), two clusters (TC), Watts-Strogatz (WS), and Barab{\'a}si-Albert (BA).}
\label{tab:parameters}
\end{table}

\subsection{Brain data}\label{sec:brain_data}
We apply \ac{SAS} to compare non-square matrices of brain activity in response to visual stimuli.
We use an openly available data set, which has an extensive description of the task and recording apparatus \citep{Chen22_77}.
In the experiments, the activity of neurons in the primary visual cortex (V1) of one macaque monkey (\textit{Macaca mulatta}) was recorded using several extracellular electrode arrays (Utah arrays, $8\times8$ electrodes).
The quality of the signals was assessed based on the signal-to-noise ratio and channel impedance.
For details of the data recording and processing we refer to \cite{Chen22_77}.
Here, we focus on a single array ($\rm ID = 11$) during a receptive field mapping task.
In this task, for each trial the macaque had to fixate its gaze to the center of the screen for $\rm 200~ms$.
Subsequently, one bright bar moved across the screen for $\rm 1000~ms$ in one of four directions: rightward (R), leftward (L), upward (U), or downward (D).
The different task modalities are in the following referred to as trial types.
For each trial type, there are $N=120$ repetitions.

The activity time series recorded from the electrodes was processed to obtain the multi-unit activity envelope (MUAe) with a sampling rate of $\rm 1~kHz$, a commonly used signal as a proxy for neuronal firing rates \citep{Super05_263}, see \citep{Morales23_bioRxiv} for specific details of the processing.
We align the trials to the peak response, defined as the maximum response from the average MUAe across electrodes, and cut data in a window  $\rm \pm 200~ms$ around the alignment trigger.
This yields one $\rm 64 \times 400$ matrix per trial: $64$ electrodes during $\rm 400~ms$ at $\rm 1~kHz$; see examples in \autoref{fig:brain_activity}B.
We then group the matrices by trial type for comparison by \ac{SAS}.

\subsection{Degenerate matrices}
\label{sec:degenerate_matrices}

To assess \ac{SAS} in the presence of degenerate singular values we construct matrices with various levels of degeneracy from random matrices and random graphs. Given a matrix $M = U \Sigma V^T$, we introduce a degeneracy parameter $d$ forcing
\begin{equation}
    \sigma^{i} = ... = \sigma^{i+d} = \frac{1}{d} \sum_{j=i}^{i+d} \sigma^{j}
\end{equation}
and define the degenerate matrix as
\begin{equation}
    M^d = U \Sigma^d V^T~,
\end{equation}
where in $\Sigma^d$ the corresponding singular values are replaced by their empirical mean.
We distinguish three special cases of starting index $i$ such that either the highest (first), lowest (last), or central singular value are first made degenerate, referred to as \textit{H-}, \textit{L-}, and \textit{C-degenerate}, respectively. In the C-degenerate case, $i$ is adjusted such that the central singular value lies in the center of the interval $[i, i+d]$. Finally, we denote the relative degeneracy of an $N \times N$ matrix as $D = d / N$.

\section{Results}\label{sec:results}
We present a measure for assessing the structural similarity between two arbitrary, real $m \times n$ matrices $M_a, M_b$ named \textit{singular angle similarity} (SAS). The measure is based on singular value decomposition (SVD), which introduces the left and right singular vectors with corresponding singular values (\autoref{eq:SVD_eq}). 
\ac{SAS} exhibits the following properties:
\begin{itemize}
    \item \ac{SAS} attains values between $0$ and $1$ where higher values imply greater similarity.
    \item \ac{SAS} is invariant under actions of identical orthogonal maps from the left or the right on the compared matrices; this includes the consistent permutation of rows and columns as a special case (\autoref{supp:invariance_props}).
    \item \ac{SAS} is invariant under transposition of both matrices (\autoref{supp:invariance_props}).
    \item \ac{SAS} is invariant under scaling with a positive factor; in particular, \ac{SAS} $= 1$ for $M_b = c_{1} M_a, c_{1} \in \mathbb{R}^+$ (\autoref{supp:scaling}).
    \item \ac{SAS} is zero if the two compared matrices are equal up to a negative factor: $\mathrm{SAS}~=~0$ for $M_b = c_{2} M_a, c_{2} \in \mathbb{R}^-$ (\autoref{supp:scaling}).
    \item if $w(x,y) = \sqrt{x \cdot y}$, SAS is invariant to isotropic scaling \citep{Kornblith19_3519}
\end{itemize}
Thus, \ac{SAS} predominantly highlights structural differences between the matrices. The derivation of this measure is presented in \autoref{sec:similarity}.

\subsection{Geometric interpretation of SAS}
\label{sec:geometric_interpretation}
\begin{figure}[h]
    \begin{center}
        \includegraphics[width=\linewidth]{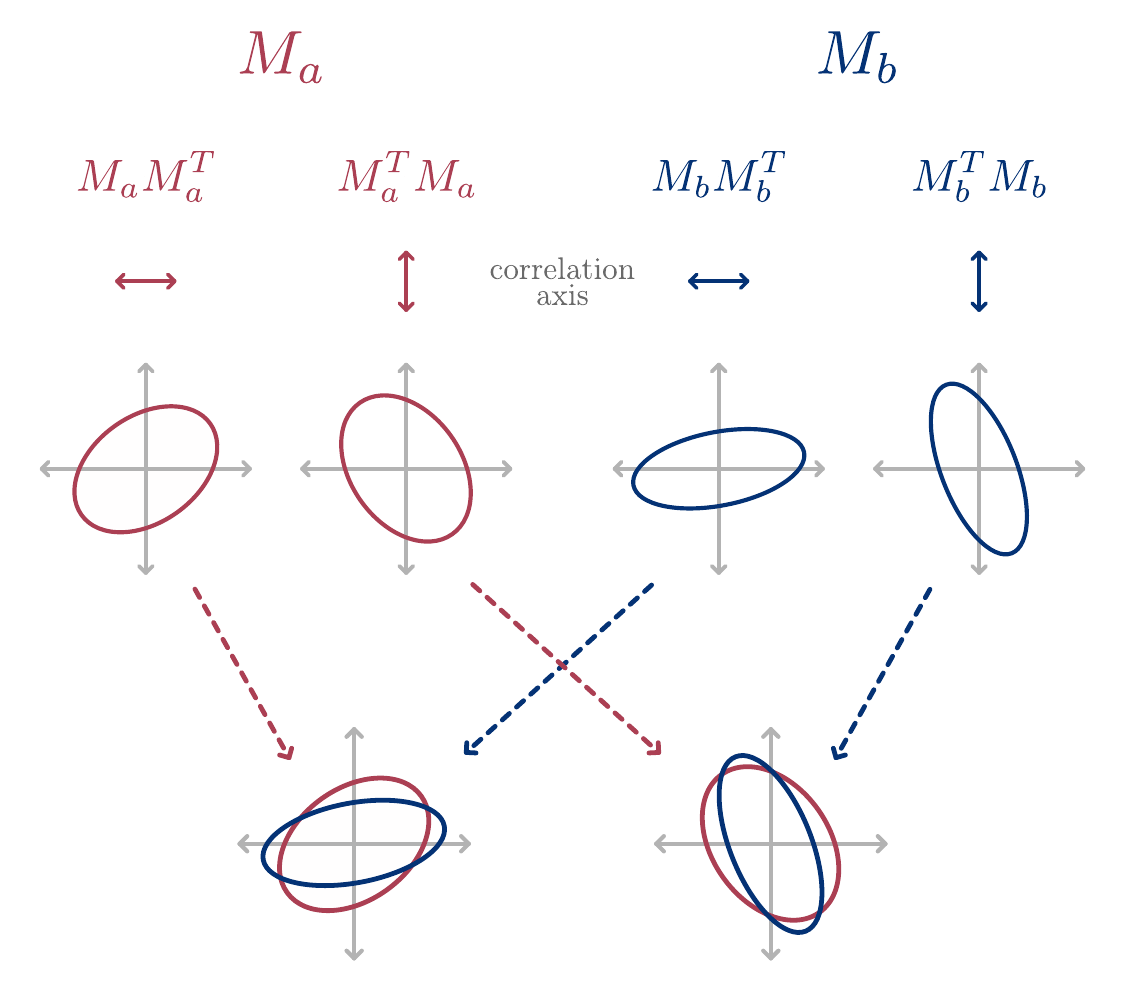}
    \end{center}
    \caption{
    \textbf{Geometric interpretation of \ac{SAS}.}
   Matrices $M_a$ (red) and $M_b$ (blue) as in \autoref{fig:erklaerbaer}. The eigenvectors of $M_{i}M_{i}^{T}$ and $M_{i}^{T}M_{i}$  (same colors) scaled by the square root of their eigenvalues span the main axes of ellipsoids. These square matrices capture the correlation structure of $M_i$ along the horizontal and vertical axis, respectively (double-headed colored arrows). \ac{SAS} compares the angles between the corresponding ellipsoids (dashed colored arrows).
    }
    \label{fig:ellipsoids}
\end{figure}

Singular angle similarity has a geometric interpretation. The left and right singular vectors of $M_i$ ($i \in \{a, b\}$) are the respective eigenvectors of the square matrices $M_{i}M_{i}^{T}$ and $M_{i}^{T}M_{i}$. Further, $M_i M_i^T$ and $M_i^T M_i$ have the same eigenvalues (the squared singular values of $M_i$). Consider for each of these symmetric positive-definite matrices a hyperellipsoid spanned by the respective eigenvectors scaled by their eigenvalues (\autoref{fig:ellipsoids}).

The hyperellipsoid collapses in most dimensions as matrices $M_i$ typically have only a small number of large singular values (c.f. \cite{Marchenko67_507}). Dimensions associated with the largest singular values dominate its shape, and the angle between the corresponding left and right singular vectors of the matrices $M_a$ and $M_b$ are of main relevance for \ac{SAS}. Thus, a high \ac{SAS} indicates that the hyperellipsoids are aligned, whereas a low \ac{SAS} indicates misalignment or different shapes. If two matrices share two-dimensional structural features, their hyperellipsoids will be similarly shaped and point into similar directions, producing a high \ac{SAS}. $M_{i}M_{i}^{T}$ and $M_{i}^{T}M_{i}$ are the correlation matrices up to a normalization by the number of rows and columns, respectively, and the subtraction of the mean. Therefore, \ac{SAS} takes into account the correlation structure along both axes of the matrices. This distinguishes the measure from common methods such as cosine similarity and the Frobenius norm.

\subsection{Comparison with standard measures for random matrices}
\label{sec:comparison_other_metrics}

\begin{figure}[!h]
    \begin{center}
        \includegraphics[width=\linewidth]{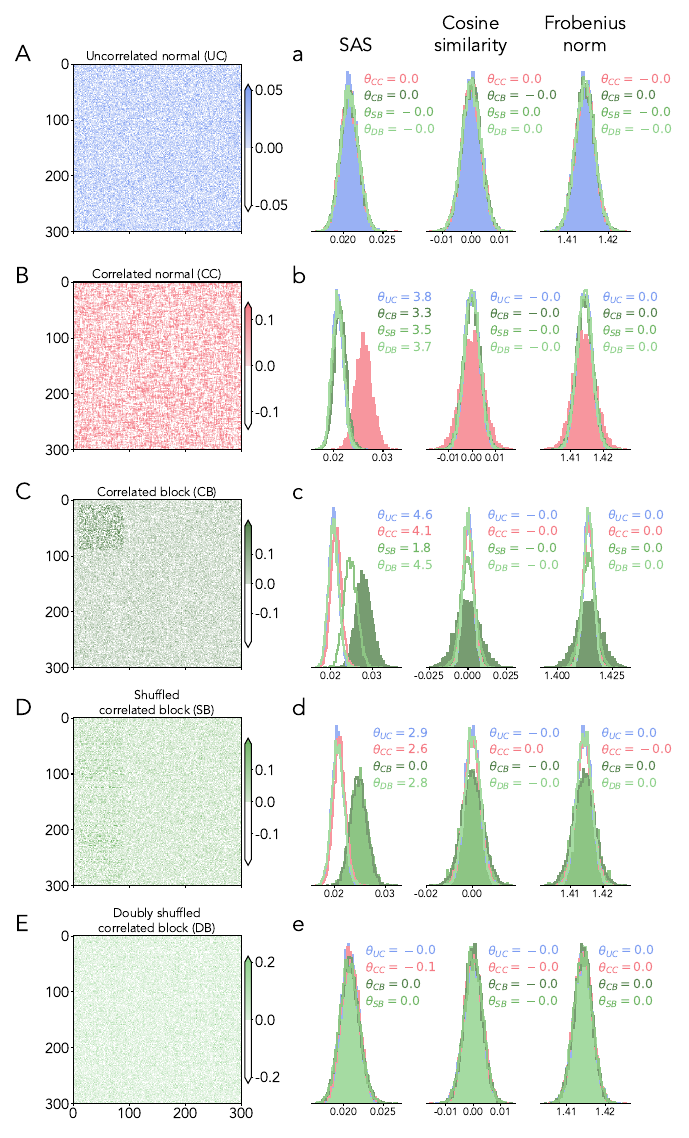}
    \end{center}
    \caption{
    \textbf{Comparison of \ac{SAS} with standard measures.}
    \textbf{A--E} Single instances of the different random matrix classes. For visibility, negative values are shown as white. 
    \textbf{a--e} Histograms of \ac{SAS}, cosine similarity, and the Frobenius norm between instances of the random matrix classes ($n=100$, all pairs compared).
    Filled distributions indicate self-similarities (self-distances), non-filled ones indicate cross-similarities (cross-distances). Legends show effect sizes $\theta$ for the comparison between distributions. 
    }
    \label{fig:comparison}
\end{figure}

By its very definition, \ac{SAS} captures two-dimensional structures that are invisible to traditional measures of matrix similarity. \autoref{fig:comparison} shows the ability of different measures to discriminate between classes of random matrices with such structure. 

Concretely, we compare \ac{SAS} with cosine similarity,
\begin{equation}
    \langle M_a, M_b \rangle_F = \mathrm{tr}(M_a M_b^T)~,
\end{equation}
and Frobenius distance,
\begin{equation}
    ||M_a - M_b||_F = \sqrt{\langle M_a - M_b, M_a - M_b \rangle_F} ,
\end{equation}
where we normalize the matrices such that
\begin{equation*}
    ||M_a||_F~= ||M_b||_F = 1~.
\end{equation*}
\autoref{fig:comparison}A--E show single instances of the five matrix classes defined in \autoref{sec:random_matrices}.
We first calculate the \textit{self-similarity}, the pairwise similarity between instances of the same random matrix class (excluding self comparisons), and the \textit{cross-similarity}, which refers to the similarity between instances of different classes (\autoref{fig:comparison}a--e). We analogously define self- and cross-distance for the Frobenius distance. Subsequently, we investigate whether the different measures distinguish the random matrix classes from each other based on realizations of their particular structures. Fundamentally, this only works if the structure quantified by a measure is more similar between instances of the same class than across classes. Thus, for \ac{SAS} and cosine similarity, the self-similarity must be meaningfully greater than the cross-similarities. Conversely, since the Frobenius norm measures a distance rather than a similarity, the self-distance must be smaller than the cross-distance. We call a difference meaningful if the effect size $\theta$ between pairs of distributions is greater than one. Assuming an underlying Gaussian model for the distributions, we employ the definition
\begin{equation}
    \theta = \frac{\mu_\mathrm{self} - \mu_\mathrm{cross}}{\sqrt{\frac{\sigma_\mathrm{self}^2 + \sigma_\mathrm{cross}^2}{2}}}
\label{eq:effect_size}
\end{equation}
of the "Cohen's D" effect size \cite{Cohen88} underlying the common Student's and Welch's t-statistics \cite{Student1908, Welch47_28}, where $\mu$ and $\sigma$ are the mean and standard deviation of the self- and cross-similarity distributions. Thus, two distributions are meaningfully different if the distance of their means is greater than the quadratic mean of their standard deviations.

\autoref{fig:comparison}a shows that UC matrices cannot be distinguished from the other matrix classes by any measure. This is expected: since the entries are independent, there is no detectable structure. In particular, this means that no structure is shared between different UC matrices or between UC matrices and matrices of other classes. Geometrically, this corresponds to ellipsoids that are oriented in random directions for each instance.

By definition, CC matrices exhibit shared fluctuations that induce similarity between different instances of the matrices. However, cosine similarity and the Frobenius norm fail to identify the common correlation structure (\autoref{fig:comparison}b). Only \ac{SAS} separates the self- and cross-similarity meaningfully and can thereby distinguish this matrix class from the others.

A similar conclusion holds true for CB matrices, where the correlated structure is embedded into an otherwise uncorrelated matrix (\autoref{fig:comparison}c): again, only \ac{SAS} separates the self- and cross-similarities meaningfully.

Next, we consider SB matrices. By construction, these are CB matrices with permuted rows. Between SB matrices, the CB correlation structure along the horizontal axis (quantified by $M M^{T}$) is destroyed while the correlation structure along the vertical axis (quantified by $M^{T}M$) stays the same. Since \ac{SAS} takes into account both, it detects similarity between SB and CB matrices despite the permutation of the rows. This leads to a higher cross-similarity between CB and SB matrices than between CB and the other matrix classes (\autoref{fig:comparison}c). Since the block structure exhibited by CB matrices can be viewed as one specific permutation of the rows, the self-similarity of SB and the cross-similarity between SB and CB follow the same distribution (\autoref{fig:comparison}d) while SB matrices are separable from UC and CC matrices. The cosine similarity and the Frobenius norm fail to separate SB matrices from the other classes. The choice of the axis along which we permute is arbitrary; the results are identical if we permute columns instead of rows.

Finally, we turn to DB matrices (\autoref{fig:comparison}e). Here, starting from CB matrices, both, the columns and rows are permuted. Consequently, between DB matrices a correlation structure is neither retained along the horizontal nor along the vertical axis. Neither \ac{SAS} nor the other measures can detect similarity between DB matrices in comparison with DB matrices and matrices of the other classes. 

Why are these examples relevant? They show that \ac{SAS} captures certain two-dimensional correlation structures between instances. In contrast, the traditional measures cannot identify them. Additionally, \ac{SAS} retains similarity even after permutation along one axis\textemdash including shifts as a special case. This is relevant in practical applications, for example in the analysis of highly parallel time series: even if the time series are not aligned, \ac{SAS} exposes structural similarities. However, if both rows and columns are randomly permuted \ac{SAS} fails to identify similarity. This is expected since in this case there is no two-dimensional correlation structure between instances left for \ac{SAS} to identify.

In addition to the metrics described above, we also test \ac{SAS} against linear CKA and the Angular Procrustes Similarity \autoref{supp:rm_supp}. While for these measures CC and CB matrices can be better separated, this does not generalize to SB matrices. Moreover, the separations suffer from spurious high similarity for UC matrices, violating the expected hierarchy of similarity in multiple cases \autoref{supp:rm_supp}.

\subsection{Characterization}
\label{sec:characterization}
\subsubsection{Scale dependence and robustness}
\begin{figure}[!ht]
    \begin{center}
        \includegraphics[width=\linewidth]{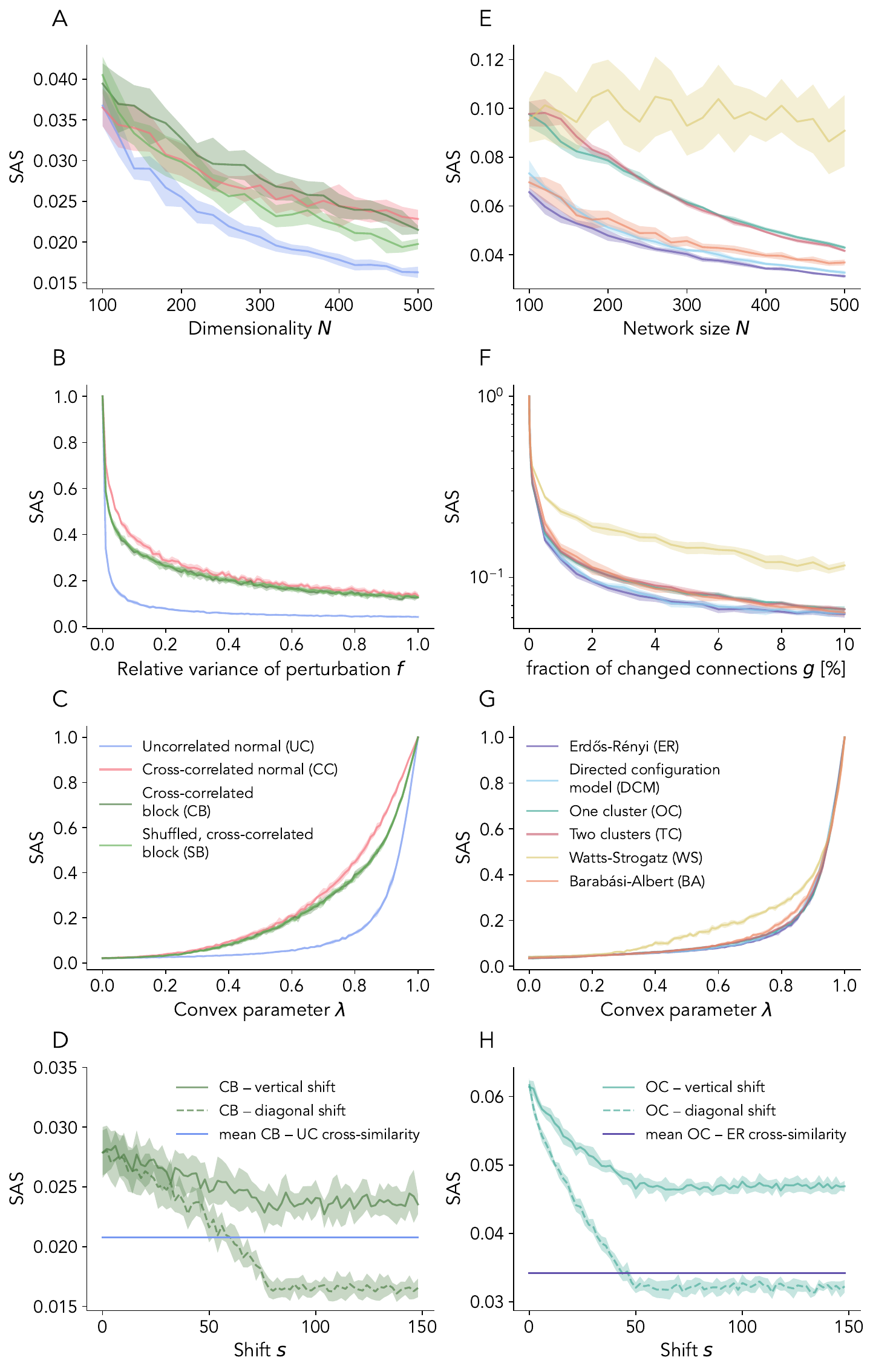}
    \end{center}
    \caption{
    \textbf{Characterization of SAS.}
    In all panels, lines indicate the mean and shadings indicate the standard deviation over 10 realizations. In cases of constant matrix size, $N=300$. A-C and E-G share the same legends.
    \textbf{A} Self-similarity for varying dimensionality $N$.
    \textbf{B} \ac{SAS} between identical matrix instances for varying variance of an additive perturbation $\sigma^2_\mathrm{pert} = f \sigma^2$.
    \textbf{C} \ac{SAS} between matrices when increasing the degree of putative structural similarity quantified by $\lambda$.
    \textbf{D} \ac{SAS} between CB matrices where the correlated block is shifted either vertically or diagonally for one matrix.
    \textbf{E} Self-similarity for varying network size $N$.
    \textbf{F} \ac{SAS} between identical network model instances where the number of individual connections that are changed is gradually increased.
    \textbf{G} \ac{SAS} between network model instances when increasing the degree of putative structural similarity quantified by $\lambda$.
    \textbf{H} \ac{SAS} between OC matrices where the cluster is shifted either vertically or diagonally for one matrix.
    }
    \label{fig:characterization}
\end{figure}

To assess the dependence of \ac{SAS} on matrix size, we calculate the self-similarity for increasing dimensionality $N$ and observe a decreasing \ac{SAS} for all random matrix classes (\autoref{fig:characterization}A). Since the probability distribution for an angle between two random vectors increasingly centers around $\pi/2$ with increasing dimensionality \citep{Cai13_1837}, the resulting \ac{SAS} between UC matrices decreases with increasing matrix dimensionality. This intuition generalizes to the other matrix classes. Hence, a quantitative comparison of \ac{SAS} values is only reasonable for matrices of the same dimensionality.

Next, we investigate how \ac{SAS} decreases between two identical copies of a matrix when gradually perturbing one of them. We analytically study \ac{SAS} between a matrix $M$ and a perturbed version of itself, $M + \sqrt{\epsilon}W$, using Rayleigh-Schr{\"o}dinger perturbation theory \citep{Landau_QM}. Here, $\sqrt{\epsilon}$ is chosen as a perturbation parameter since this ensures a linear scaling of the variance of the perturbation matrix with $\epsilon$. We find that, for a large class of perturbations, \ac{SAS} follows $1 - \arccos\left(1 - O(\epsilon)\right) / \frac{\pi}{2}$ (see \autoref{supp:perturbation}). This implies that small differences are identified as dissimilarities arbitrarily fast ($\frac{d\arccos(1-x)}{dx}~\rightarrow~\infty$ for $x~\rightarrow~0$). Thus, \ac{SAS} is sensitive to small differences in the compared matrices. For an empirical analysis, we study the sensitivity of \ac{SAS} under perturbations of additive noise of the form $\tilde{M} = M + W$ where $W$ is a matrix of the UC class with zero mean and variance $\sigma^2_{\mathrm{pert}} = f \sigma^2$, $0 \leq f \leq 1$. As predicted analytically, \autoref{fig:characterization}B shows a rapid fall-off for small perturbations that continues as a gradual decrease for each of the considered classes.

Next, we numerically study \ac{SAS} while adding structure to a noise matrix. In particular, we calculate \ac{SAS} between a matrix $M$ and the convex combination of the same matrix with a UC matrix $N$:
\begin{equation}
    (1 - \lambda) N + \lambda M \qquad \textnormal{for } \lambda \in [0, 1]~.
\end{equation}
\autoref{fig:characterization}C shows that \ac{SAS} is well-behaved also for small values of $\lambda$: it increases smoothly when adding structure for all matrix classes.

Finally, starting from CB matrices, we investigate the change of cross-similarity when shifting the correlated block of one matrix either vertically or diagonally by $s$ indices (\autoref{fig:characterization}D). In both cases, \ac{SAS} gradually decreases and saturates once the blocks are non-overlapping ($s=80)$. Importantly, SAS saturates to a value higher than the mean CB – UC cross-similarity for the vertical shift, whereas it saturates to a value lower than that for the diagonal shift. Thus, SAS identifies shared structure when it is shifted vertically, but not when it is shifted diagonally. While potentially counterintuitive, this can be understood when considering the case of DB matrices (\autoref{fig:comparison}e): the diagonal shift is a specific example of a permutation in both directions, and once the blocks are non-overlapping, there is no correlation structure between the two matrices that SAS can identify. We conclude that SAS cannot detect similarity between matrices if a sufficient amount of the relevant structure is moved across instances.

We perform an analogous analysis for network adjacency matrices of six different graph models: Erd\H{o}s-R{\'e}nyi (ER), directed configuration model (DCM), one cluster (OC), two clusters (TC), Watts-Strogatz (WS), and Barab{\'a}si-Albert (BA), as defined in \autoref{sec:random_graphs}. The results are qualitatively similar to those obtained for the four classes of random matrices: \ac{SAS} decreases with increasing network size $N$ (\autoref{fig:characterization}E), \ac{SAS} rapidly decreases for small perturbations (\autoref{fig:characterization}F), \ac{SAS} gradually increases when adding structure (\autoref{fig:characterization}G), and \ac{SAS} decreases when shifting either vertically or diagonally (\autoref{fig:characterization}H). A notable exception is that for WS networks, \ac{SAS} does not decrease when increasing $N$. In this network model the number of nearest neighbors each node is connected to scales with $N$. Therefore the correlation in the adjacency matrix also scales with $N$, rendering the similarity measured by \ac{SAS} independent of $N$. For investigating the effect of a perturbation on identical network matrices (\autoref{fig:characterization}F), we define the gradual change such that an increasing fraction $g$ of matrix elements is altered. Specifically, for each of the $g N^2$ randomly selected matrix elements, existing connections are removed and missing connections are established with a multiplicity of one. For the binary adjacency matrices this corresponds to bit-flipping the corresponding entries. In the case of adding structure (\autoref{fig:characterization}G), we choose the ER adjacency matrices as the noise component $N$. Note that while the sum over all entries stay the same on average under the convex combination, the entries are not confined to natural numbers anymore.

\subsubsection{Degenerate singular values}

\begin{figure}[!ht]
    \begin{center}
        \includegraphics[width=\linewidth]{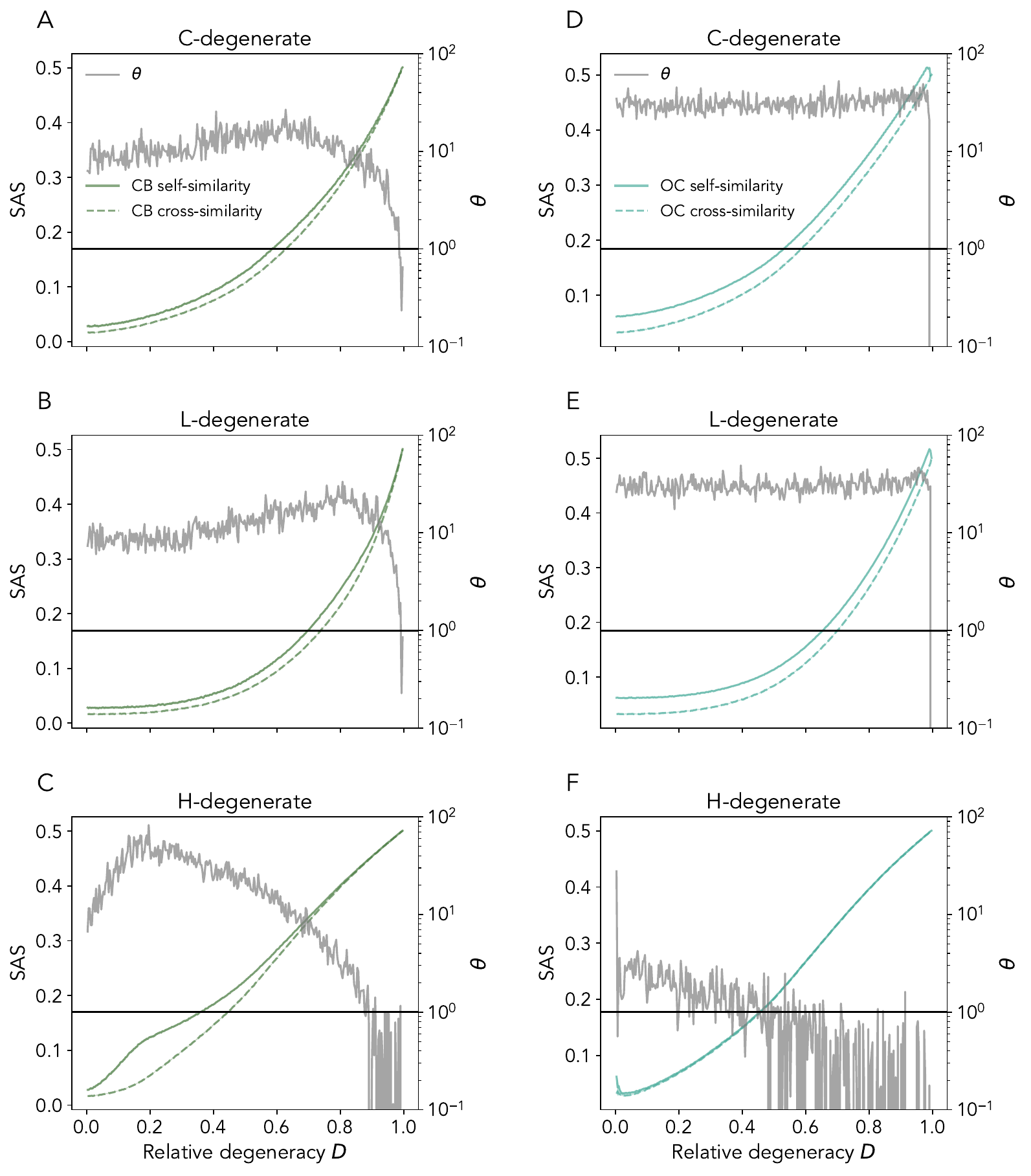}
    \end{center}
    \caption{
    \textbf{SAS for degenerate matrices.}
    In all panels, lines indicate the mean, the shading the standard deviation over 10 realizations. Solid colored lines show self-similarity between CB (OC) matrices with identical parameters (same as in \autoref{tab:parameters_matrices} and \autoref{tab:parameters}). Dashed colored lines show cross-similarity between two CB (OC) matrices where the second matrix exhibits a block (cluster) shifted to $b^{\mathrm{lower}} = 210$ and $b^{\mathrm{upper}} = 290$ ($b^{\mathrm{lower}} = 200$ and $b^{\mathrm{upper}} = 250$). Gray lines indicate effect size $\theta$ (right axes). In all panels, the fraction $D$ of degenerate singular values is varied. $N=300$. 
    \textbf{A,D} C-degenerate matrices, where the singular values are made degenerate starting from the center singular value. 
    \textbf{B,E} L-degenerate matrices, where the singular values are made degenerate starting from the lowest singular value. 
    \textbf{C,F} H-degenerate matrices, where the singular values are made degenerate starting from the highest singular value. 
    }
    \label{fig:degenerate_singular_values}
\end{figure}
We evaluate the discriminability of matrix classes with SAS in the presence of degenerate singular values (\autoref{fig:degenerate_singular_values}).
Specifically, we quantify how well CB matrices with non-overlapping blocks and OC matrices with non-overlapping clusters can be discriminated when making the singular value spectrum degenerate. We introduce degeneracy gradually as described in \autoref{sec:degenerate_matrices}.

In the case of CB matrices, for which the singular value spectrum exhibits a smooth decay (not shown), SAS is able to distinguish matrices with blocks at different positions even for high degeneracy ($\theta > 1$ up to $D \approx 0.85$) as shown in \autoref{fig:degenerate_singular_values}A--C. Here, the effect size initially increases with increasing H-degeneracy even though the destroyed structure corresponds to the singular angles with the highest weighting in the calculation of SAS.

In contrast, the spectrum of singular values of OC matrices is dominated by one singular value (not shown), which is of highest relevance for \ac{SAS}. When introducing degeneracy from the center (\autoref{fig:degenerate_singular_values}D) or starting from the smallest singular value (\autoref{fig:degenerate_singular_values}E), $\theta$ remains roughly constant until $d=299$, $d=300$, respectively. At precisely these values, the largest singular value becomes degenerate and $\theta$ drops below $1$. For the left-degenerate case (\autoref{fig:degenerate_singular_values}F), some discriminability is retained for small values of $d$ even though the largest singular value is made degenerate already at $d=2$.

This implies that, in a certain range depending on the use case, SAS is robust against the perturbation of large outliers in the distribution of singular values. Moreover, these cases suggest that SAS is not principally limited by the presence of repeated singular values.
\subsection{Categorization of random graphs}
\label{sec:categorization_random_graphs}

\begin{figure*}[ht]
    \begin{center}
        \includegraphics[width=\textwidth]{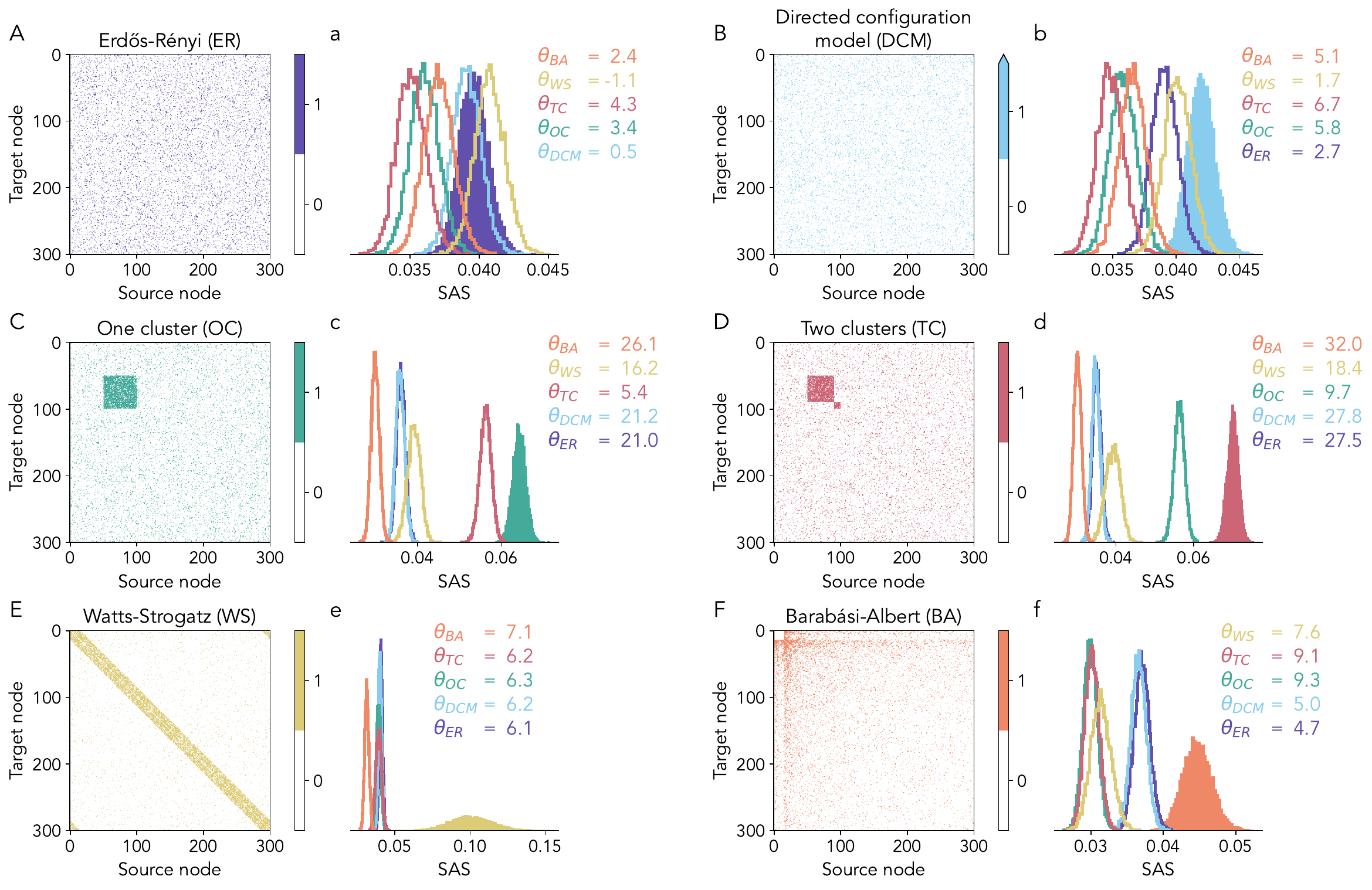}
    \end{center}
    \caption{
    \textbf{Self- and cross-similarity of different network models.} 
    \textbf{A--F} Single instances of the different network models.
    Colored matrix elements indicate a connection between nodes. For the DCM model, connections with a multiplicity higher than one are shown with the same color intensity as single connections.
    \textbf{a--f} Histograms of \ac{SAS} between instances of the network models ($n=100$, all pairs compared). Filled distributions indicates self-similarities, non-filled ones indicate cross-similarities. Legends show effect sizes $\theta$ for the comparison between self- and cross-similarities.}
\label{fig:connectomes_and_similarity}
\end{figure*}
We apply \ac{SAS} to the adjacency matrices that describe the network architecture in various directed and undirected probabilistic graphs as defined in \autoref{sec:random_graphs}. Example adjacency matrices of network instances are shown in \autoref{fig:connectomes_and_similarity}A--F. Since these matrices $M_i$ only contain non-negative entries, so do $M_i M_i^{T}$ and $M_i^{T} M_i$. The Perron-Frobenius theorem \citep{Deuflhard08} guarantees that the left and right singular vectors corresponding to the largest singular value have only non-negative or only non-positive entries (cf. \autoref{supp:perron}). As such, these singular vectors are confined to a single orthant of the $N$-dimensional vector space. Even if these vectors are random, they cannot be assumed to be orthogonal. Indeed, for the ER network, for which all other singular vectors are of random orientation, the first left and right singular vectors scatter around the vector $\left( 1 / \sqrt{N}, ..., 1 / \sqrt{N} \right)^T$ across instances. Therefore, the first singular vectors necessarily enclose smaller angles across models compared to the other pairs of singular vectors. Consequently, information regarding the difference between models\textemdash which is encoded most strongly in the first singular vectors\textemdash is reduced. Thus, it is \textit{a priori} not clear whether \ac{SAS} reliably distinguishes between network models.

\paragraph{Self-similarity of network models}
First, we examine the self-similarity of the network models (\autoref{fig:connectomes_and_similarity}a--f). We find that \ac{ER} networks exhibit the lowest self-similarity compared to all other network models. This is consistent with the \ac{ER} network model maximizing the entropy under the constraint that the average number of connections is constant: \ac{ER} networks have the least structure that is shared across instances. This can be also understood from their definition inasmuch as each connection is realized independently with the same probability. In this sense, ER networks are analogous to the UC random matrices. The other network models instead feature structural properties that are consistent across instances, stemming from shared variations in the connection probability. This is most obvious for the OC and TC network models (analogous to the CB random matrix class), where certain subgroups of nodes have a higher connection probability $p$ among themselves compared to the rest of the network. Further, we expect WS networks to have reliably detectable structure, i.e., high self-similarity, as every node has dominant local connectivity. \ac{SAS} confirms these expectations, as seen when comparing the respective self-similarity distributions in \autoref{fig:connectomes_and_similarity}a--f.

\paragraph{Self-similarity vs. cross-similarity}
Second, we study whether \ac{SAS} can differentiate between the particular structures present in the adjacency matrices of the network models.
Using the effect size defined in \autoref{eq:effect_size}, \autoref{fig:connectomes_and_similarity}a--f confirm that the self-similarity is meaningfully higher than the cross-similarity except for special cases.

First, \ac{ER} networks do not exhibit $\theta > 1$ for all cases.
As for the UC matrices, this is expected as all matrix elements are uncorrelated.
As a matter of fact, the cross-similarity with WS networks yields even higher \ac{SAS} values than the self-similarity of ER.
Why is this the case?
The first left and right singular vectors of both ER and WS networks scatter around $\left( 1 / \sqrt{N}, ..., 1 / \sqrt{N} \right)^T$.
The deviation between singular vectors of ER networks, however, is larger than that between those of WS networks across instances.
This leads to a better alignment, i.e., a higher angular similarity, of the singular vectors, resulting in a higher \ac{SAS} between ER and WS as compared to ER and ER networks.

Second, we note that \ac{SAS} distinguishes between the OC and TC networks despite overlapping clusters.
\autoref{fig:connectomes_and_similarity}c--d show that the respective self-similarities are closer to the cross-similarity of OC and TC than to the other cross-similarities. Thus, \ac{SAS} identifies the clustered networks to be more similar among each other than compared to the remaining networks.

We conclude that \ac{SAS} is sensitive to the structure present in matrices, enabling it to distinguish between model classes.
The same conclusion also holds true for non-square matrices of network connectivity where a full graph is instantiated, but only subsamples are analyzed with \ac{SAS} (see \autoref{supp:non-square-networks}).

\subsection{Separation of brain states}
\label{sec:separation_brain_states}

\begin{figure*}[!h]
    \begin{center}
        \includegraphics[width=\textwidth]{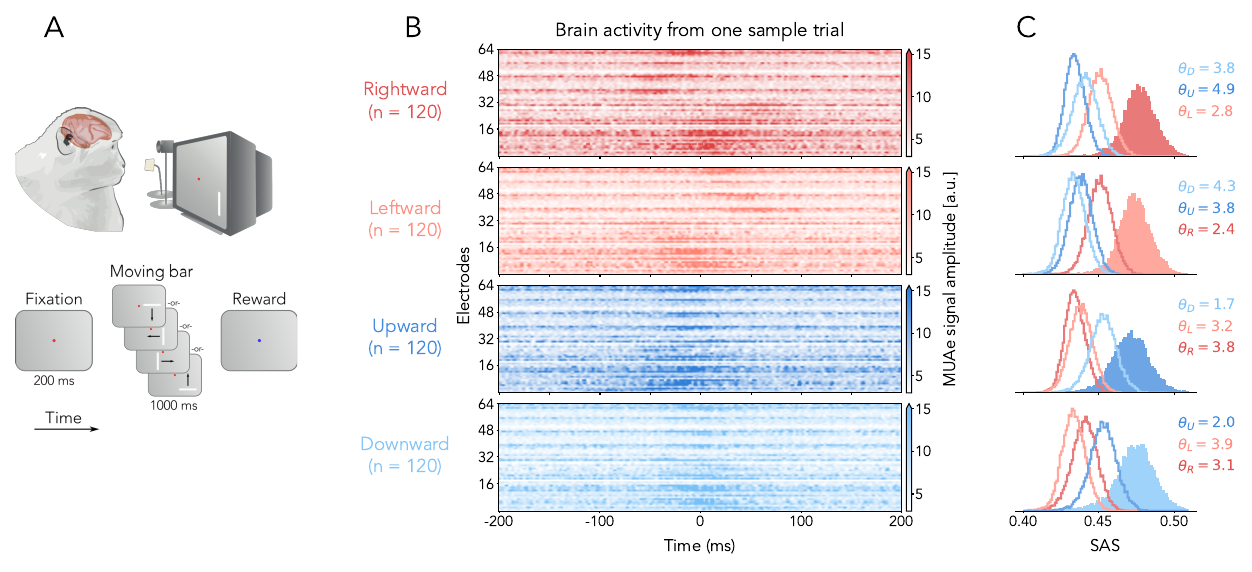}
    \end{center}
    \caption{
    \textbf{Comparing non-square matrices of brain activity with \ac{SAS}.} 
    \textbf{A} Schematic diagram of the neuroscience experiment. 
    \textbf{B} Single-trial brain activity at identical recording electrodes ($64$ electrodes, vertical) for each type as example (trial types distinguished by color). Larger values of matrix entries indicate higher MUAe activity (shading in color bar, arbitrary units).
    \textbf{C} Histograms of \ac{SAS} between all ($n=120$) individual trials. Filled distributions (color code as in B) indicate self-similarities, non-filled ones indicate cross-similarities. Legends show effect sizes $\theta$ for the comparison between self- and cross-similarities.}
    \label{fig:brain_activity}
\end{figure*}

We investigate brain activity originating from different experimental trials as a use case for \ac{SAS} with non-square matrices of experimental data. The publicly available data set from \cite{Chen22_77} is based on extracellular recordings from the visual cortex of a macaque monkey. In the experiment, bright bars move across a screen in one of four directions (rightward (R), leftward (L), upward (U) or downward (D)), evoking a strong neural response (\autoref{fig:brain_activity}A). The data consists of the multi-unit activity envelope (see \autoref{sec:methods}) yielding one $64\times400$ matrix per trial; sample matrices are shown in \autoref{fig:brain_activity}B.

Neurons in the primary visual cortex (V1) respond according to their feature selectivity, primarily stimulus location \citep{Tootell82_902} and orientation \citep{Hubel59}, but also movement direction \cite{DeValois82}. Beyond these well-known response properties of single neurons, the population activity\textemdash represented as a two-dimensional spatio-temporal matrix\textemdash may reveal additional information about brain dynamics. By applying \ac{SAS}, we investigate shared variability across both time and neurons.

In the data set at hand, \ac{SAS} reveals that neural activity of all trial types exhibits higher self- than cross-similarity with effect sizes $\theta>1$ (\autoref{fig:brain_activity}C). Trials with stimulus movement along the same axis (L-R or U-D) are more similar to each other than to ones with orthogonal stimulus movement. This is a desirable outcome of \ac{SAS}: in L-R (resp. U-D) trials, neurons that share an orientation tuning aligned with the stimulus are expected to have a higher probability of a strong response. The consideration implies that the shared orientation of the bar stimulus in L and R (resp. U and D) trials leads to more similar responses in these trial pairs.

In this use case, \ac{SAS} outperforms common measures of matrix similarity. Both cosine similarity and the Frobenius norm can distinguish trial types\textemdash albeit with lower $|\theta|$ than \ac{SAS}\textemdash but fail to identify the shared variability along the same axis (L-R or U-D). Symmetric CKA and the Angular Procrustes Similarity fail to reliable distinguish trials in most cases (\autoref{supp:brain_activity}). Therefore, the use case highlights the ability of \ac{SAS} to identify the two-dimensional structure of matrices in experimental data.

As an essentially linear measure operating directly on the data, SAS might suffer in the presence of strong non-stationarities and highly non-linear trajectories in the phase spaces. \citet{Ostrow23_33824} suggest to overcome this problem by representing the non-linear system as a high-dimensional linear one using dynamic mode decomposition \citep{Brunton21}. In a second step, they then assess the similarities of the linear representations via a Procrustes-style minimization problem using base-changes instead of actions of orthogonal matrices. \citet{Guilhot24} show that the method can identify the emergence of representations during learning in recurrent networks \textemdash a task at which static measures like CKA or a Procrustes based approach fail. It will be interesting to see how the assessment of similarity changes if SAS is used instead of the similarity measure employed by the authors in \citep{Ostrow23_33824} for comparing the linear high-dimensional representations of the dynamical systems.

\section{Discussion}\label{sec:discussion}
Here, we present \textit{singular angle similarity} (\ac{SAS}): a method for comparing the similarity of real matrices of identical shape based on singular value decomposition. \ac{SAS} is invariant under transposition and consistent relabeling of coordinates, and reliably detects structural similarities in the compared matrices. It generalizes the angular similarity (\autoref{supp:1d_case}) for vectors and the eigenangle score for symmetric positive definite matrices by \citet{Gutzen23_104813}. For a particular choice of weight function and vector similarity measure, \ac{SAS} and the eigenangle score are equivalent for positive-definite symmetric matrices.
\citeauthor{Gutzen23_104813} introduce an extension of the eigenangle score for asymmetric adjacency matrices of certain network types by choosing a specific analytical mapping of the complex-valued eigenvalues and vectors: shift the real part of the eigenvalues by the spectral radius of the matrix, and calculate the Euclidean angle between the complex eigenvectors \cite{Scharnhorst01_95}. However, this choice is not unique\textemdash other mappings are also possible. In addition, the resulting similarity measure is not invariant under joint transposition of the compared matrices. In \ac{SAS}, the singular vectors are real-valued, the singular values are non-negative, and the measure is transposition invariant. Thereby, it circumvents these limitations and admits a more natural generalization to non-symmetric and non-square matrices. We here choose angular similarity as the vector similarity measure, and a specific weight function for the resulting angles (\autoref{eq:weight_function}). Other choices are possible, such as cosine similarity for the former, making \ac{SAS} easily adaptable. By definition, SAS can only be applied to matrices with identical shape. A possible extension that allows one to apply SAS to matrices of different shape is sketched in \autoref{sec:diff_shapes}.

\paragraph{Interpretability}
Since \ac{SAS} is sensitive to matrix size (\autoref{sec:characterization}) it can only be interpreted in relative terms. A quantitative reference can be obtained by choosing a use-case-specific null model (e.g. Erd\H{o}s-R{\'e}nyi for network models) for which the corresponding distribution of \ac{SAS} can be determined numerically. Similarity as indicated by \ac{SAS} can then be interpreted in terms of this baseline. Since realizations of single matrices exhibit fluctuations, evaluating \ac{SAS} from a single observation may be misleading. Instead, one should consider the \ac{SAS} distribution of an ensemble of realizations when possible. Such a distribution can then be interpreted with respect to the reference distribution obtained with the null model by means of an effect size (\autoref{eq:effect_size}) or a statistical two-sample test of choice. Given the broad range of potential use cases, analytical descriptions of a null-distribution and associated calculations of $p$-values as in \citep{Gutzen23_104813} seem difficult or even impossible for many cases, and we suggest an empirical approach based on explicit choices of null models as described above.

However, we highlight that an empirical approach to a statistical assessment of methods based on the analyses of matrices is not always required. Indeed, some methods for which key quantities are related to singular values or eigenvalues allow for analytical descriptions of statistical tests or other assessments (e.g. \citep{Kritchman08_19, Kritchman09_3930, Veraart16_1582, Safavi21_1751, Safavi23_e1010983}). Even though the overlap between eigenvectors has also been investigated analytically in special cases \citep{Bun18_052145}, focusing on singular angles makes a fully mathematical derivation of significance tests in many cases impossible to the best of our knowledge. We hence advocate for the empirical approach outlined above.

\paragraph{Limitations}
While \ac{SAS} generally highlights structural features in matrices stemming from their correlations along rows and columns, it also suffers from shortcomings in certain situations. In the presence of sufficiently strong noise, the order of singular values may change even if two matrices encode the same underlying information. This leads to a different pairing of singular vectors when computing \ac{SAS}, resulting in a low score even though the matrices result from a common construction process. This can be partially be addressed by rounding singular values, thereby potentially introducing degeneracy. This is treated in SAS by deriving singular angles from the canonical angles between the degenerate subspaces. If the degenerate subspaces are large, SAS may lose its sensitivity: in the extreme case of fully degenerate matrices where the degenerate subspace is the full space, no discrimination is possible using SAS.
Additionally, if the matrices under consideration have few strongly dominating singular values, and these are degenerate, the discriminability using SAS may suffer.
Further, SAS cannot detect similarity between matrices if rows and columns are inconsistently permuted. This includes shifts along both axes as a special case. Hence, SAS cannot be used when translation invariance is desired, for instance in the detection of objects in images independent of their location.
\paragraph{Applications}
Beyond network connectivity or brain activity matrices, potential applications include the analysis of non-symmetric matrices obtained with measures for the flow of information. Examples of such measures include, but are not limited to Granger Causality \citep{Granger69_424}, or Transfer Entropy \citep{Schreiber00_461} (also see \citep{Peters17_causal_inference} for a machine learning inspired view on the topic). Additionally, \ac{SAS} can help assess similarity in more classical settings, e.g., when studying cross-covariances.\newline

In conclusion, \ac{SAS} can be used to analyze the structural similarity of any real-valued data that can be represented in matrix form. Such data can come from any field of research. Coupled with domain knowledge, \ac{SAS} may reveal hidden structures in the data, supporting existing methodologies and enabling new insights.

\subsection*{Code and data availability}
Code to calculate singular angle similarity (\ac{SAS}) is openly available on GitHub (\href{https://github.com/INM-6/SAS}{https://github.com/INM-6/SAS}) and Zenodo (\href{https://doi.org/10.5281/zenodo.10680478}{https://doi.org/10.5281/zenodo.10680478}), and in the validation test library NetworkUnit\footnote{\href{https://github.com/INM-6/NetworkUnit}{https://github.com/INM-6/NetworkUnit}, RRID:SCR\_016543} \citep{Gutzen18_90}. The data and code to reproduce the results from this paper can be found on Zenodo (\href{https://doi.org/10.5281/zenodo.10680810}{https://doi.org/10.5281/ zenodo.10680810}).

\subsection*{Acknowledgments}
This work has been supported by NeuroSys as part of the initiative “Clusters4Future” by the Federal Ministry of Education and Research BMBF (03ZU1106CB); the DFG Priority Program (SPP 2041 "Computational Connectomics");
the EU's Horizon 2020 Framework Grant Agreement No. 945539 (Human Brain Project SGA3);
the European Union’s Horizon Europe Programme under the Specific Grant Agreement No. 101147319 (EBRAINS 2.0 Project). 
the Ministry of Culture and Science of the State of North Rhine-Westphalia, Germany (NRW-network "iBehave", grant number: NW21-049); the Joint Lab "Supercomputing and Modeling for the Human Brain". The authors thank the two anonymous reviewers for their constructive remarks and helpful suggestions that greatly improved the quality of the manuscript.
%

\subsection*{Author contributions}
Conceptualization: JA, AK, RG; Methodology: AK, JA, RG; Software: JA, RG, AK, AMG; Formal analysis: JA, AK, RG, AMG; Writing - original draft: JA, AK, RG, AMG; Writing - review \& editing: all; Visualization: JA, AK, AMG, RG; Supervision: MDe, SvA, MDi; Funding acquisition: SG, SvA, MDi

\printbibliography

\clearpage
\onecolumn

\setcounter{section}{0}
\renewcommand{\thesection}{S\arabic{section}}
\section{Supplementary materials}

\setcounter{figure}{0} 
\renewcommand{\thefigure}{S\arabic{figure}}
\renewcommand{\theHfigure}{S\arabic{figure}}


\subsection{\texorpdfstring{Canonical angles between subspaces}{Canonical angles between subspaces}}
\label{supp:canonical_angles}
We follow \citep{Zhu13_325}. Given two subspaces $H_{a}$ and $H_{b}$ of $\mathbb{R}^{n}$ of dimension $p$ and $q$, the canonical angles $\theta_{i}$ can be iteratively defined as
\begin{equation}
    \cos(\theta_{i}) = \max_{v_{i}\in H_{a}, w_{i} \in H_{b}}\{\langle v_{i}, w_{i}\rangle \textnormal{ s.t. } ||v_{i}||=||w_{i}||=1 \textnormal{ and } v_{i} \perp v_{1}, ..., v_{i-1}, w_{i} \perp w_{1}, ..., w_{i-1} \},
\end{equation}
or equivalently
\begin{equation}
    \theta_{i} = \min_{v_{i}\in H_{a}, w_{i} \in H_{b}}\{\arccos\left(\langle v_{i}, w_{i}\rangle \right) \textnormal{ s.t. } ||v_{i}||=||w_{i}||=1 \textnormal{ and } v_{i} \perp v_{1}, ..., v_{i-1}, w_{i} \perp w_{1}, ..., w_{i-1} \}~.
\end{equation}
for $i=1,...,\min\{p,q\}$. The vectors $v_{i}$ and $w_{i}$ are called the \textit{canonical vectors} and determine the alignment of the subspaces. If $B_{a}, B_{b}$ are orthonormal bases of $H_{a}$ and $H_{b}$ respectivley (written as matrices of $n\times p$ and $n \times q$, respectively), one can determine the canonical angles by means of the \ac{SVD} of $B_{a} B_{b}^{T}$: writing
\begin{equation}
    B_{a} B_{b}^{T} =  U \Sigma V^{T}
\end{equation}
where $U \in \mathbb{R}^{p \times p}$ and $V \in \mathbb{R}^{q \times q}$ are orthogonal matrices and $\Sigma \in \mathbb{R}^{p \times q} = \mathrm{diag}(\sigma_{1},...,\sigma_{\min\{p,q\}})$ is a diagonal matrix with entries greater than or equal to zero. As in the main text we assume that the singular values are in decreasing order. The canonical angles are then given by $\theta_{i}=\arccos(\sigma_{i})$, and $v_{i}=B_{a}Ue_{i}$ as well as $w_{i}=B_{b}Ve_{i}$.

\subsection{\texorpdfstring{Invariance properties of \ac{SAS}}{Invariance properties of SAS}}
\label{supp:invariance_props}
Let $M_{a}, M_{b} \in \mathbb{R}^{m \times n}$ with \ac{SVD}s
\begin{equation}
    M_{i}= U_{i} \Sigma_{i} V_{i}^{T}
\end{equation}
for $i\in \{a, b\}$. Noting that the transposition of an orthogonal transformation is again orthogonal, the \ac{SVD}s of $M_{i}^{T}$ are
\begin{equation}
    M_{i}^{T} = V_{i} \Sigma_{i}^{T} U_{i}^{T}~.
\end{equation}
Thus, by its definition, the \ac{SAS} between $M_{a}$ and $M_{b}$ is invariant under transposition of both matrices.
If working with the eigendecomposition instead of the SVD, one obtains, in the general case,
\begin{equation}
    M_i = P_i \Lambda_i P_i^{-1}
\end{equation}
where $\Lambda_i$ is a diagonal matrix containing the eigenvalues, and the columns of $P_i$ are the eigenvectors. Here, the eigenvectors are not guaranteed to be orthogonal. Thus, 
\begin{equation}
    ({P_i^{-1}})^T \neq P_i
\end{equation}
in general. Consequently, any measure comparing the eigenvectors is not invariant under transposition of the compared matrices. Specifically, this includes the eigenangle score \cite{Gutzen23_104813}.

Additionally, \ac{SAS} is invariant under actions of identical orthogonal matrices from the left or right onto $M_{i}$.
We denote the orthogonal matrices as $O_{1}$ and $O_{2}^{T}$.
The action of the matrices from the left and from the rights yields
\begin{equation}
    \Tilde{M}_{i} = O_{1}M_{i}O_{2}^{T}~,
\end{equation}
and the corresponding \ac{SVD} is
\begin{equation}
    \Tilde{M}_{i} = \Tilde{U}_{i} \Sigma_{i} \Tilde{V}_{i}^{T} = \big(O_{1}U_{i} \big) \Sigma_{i} \big(O_{2}V_{i}\big)^{T}~.
    \label{supp:SVD_eq_perm}
\end{equation}
Note that the singular values remain identical.
Since the left and right singular vectors of the matrices $\tilde{M}_{i}$ are $\Tilde{u}_{i}^{j} = \Tilde{U}_{i}e^{j}$ and $\Tilde{v}_{i}^{j} = \Tilde{V}_{i}e^{j}$, we compute the left and right singular angles as
\begin{equation}
    \alpha^{j}(\Tilde{U}) = \arccos \left( \langle \Tilde{U}_{a}e^{j}, \Tilde{U}_{b}e^{j} \rangle \right) = 
    \arccos \left(\langle O_{1}U_{a} e^{j}, O_{1}U_{b} e^{j} \rangle \right)
    = \arccos \left(\langle U_{a} e^{j}, U_{b} e^{j} \rangle \right)
    = \alpha^{j}(U)
\end{equation}
and \textit{mutatis mutandis} for $\alpha^{j}(\Tilde{V})$, resp. $\alpha^{j}(\Tilde{U}^{-}), \alpha^{j}(\Tilde{V}^{-})$.
We here used the orthogonality of the matrices $O_{i}$. Thus, the singular angles, the singular values, and consequently the similarity score, are invariant under the action of orthogonal matrices from the left and the right. 

\subsection{\texorpdfstring{Change in \ac{SAS} under scaling}{Change in SAS under scaling}}\label{supp:scaling}
Consider a matrix $M$ with \ac{SVD}
\begin{equation}
    M = U \Sigma V^{T}
\end{equation}
and scalars $c_{1}>0$ as well as $c_{2} < 0$. In the following, our aim is to determine the \ac{SAS} between $M$ and $c_{i}M$ for $j=1,2$.
\paragraph{Positive scaling}
We note that the \ac{SVD} of $c_{1}M$ is
\begin{equation}
    c_{1}M = U (c_{1}\Sigma) V^{T}~.
\end{equation}
Hence, only the singular values are scaled while the left and right singular vectors remain constant. This implies $\alpha^{j}=0 ~\forall j$, and thus an angular similarity of $\Delta^{j}=1 ~\forall j$. Therefore, \ac{SAS} between $M$ and $c_{1}M$ equals $1$.
\paragraph{Negative scaling}
We write $c_{2}= -1 \cdot |c_{2}|$ and determine the \ac{SVD} of $c_{2}M$ as 
\begin{equation}
    c_{2}M = (-1 \cdot U) |c_{2}|\Sigma V^{T} = U |c_{2}|\Sigma(-1 \cdot V^{T})~.
\end{equation}
In the following, we focus on the first of the two representations.
The second one can be treated analogously.
We first calculate
\begin{equation}
    \alpha^{j}(U) =\arccos \left( \langle u^{j}, -u^{j}\rangle \right) = \arccos(-1) = \pi, \quad \alpha^{j}(V) = 0~.
\end{equation}
Thus, we obtain $\bar{\alpha}^{j} = \pi / 2$ and consequently $\alpha^{j} = \pi / 2 ~\forall j$. Therefore, the angular similarity $\Delta^{j} = 1 - \alpha^{j} / \frac{\pi}{2}=0$, and the \ac{SAS} between $M$ and $c_{2}M$ equals $0$.

\subsection{CKA and Angular Procrustes for Random Matrices}
\label{supp:rm_supp}
We compare SAS with two additional similarity measures: symmetric CKA and the Angular Procrustes Similarity.
The central kernel alignment (CKA, \citep{Cortes12_795, Kornblith19_3519}) with linear kernels is given by
\begin{equation*}
    CKA(M_{a}, M_{b}) = \frac{||M_{a}M_{b}^{T}||_{F}}{\sqrt{||M_{a}M_{a}^{T}||_{F}||M_{b}M_{b}^{T}||_{F}}}~.
\end{equation*}
In terms of \autoref{sec:geometric_interpretation}, CKA captures the correlations along the horizontal axis. Since correlation structures are exhibited along both axes for the random matrices investigated in this study, we define the symmetric CKA as
\begin{equation*}
    CKA^{\mathrm{sym}}(M_{a}, M_{b}) = \frac{CKA(M_{a}, M_{b}) + CKA(M_{b}, M_{a})}{2}~.
\end{equation*}
The Angular Procrustes Distance \citep{Williams21_4738} is given by
\begin{equation*}
    \min_{O \textnormal{ orthogonal}} \arccos\left( \frac{\langle M_{a}, OM_{b} \rangle_{F}}{||M_{a}||_{F}||M_{b}||_{F}} \right) = \arccos \left(\sum_{j} \sigma^{j} \right) = \phi
\end{equation*}
where $\sigma^{1},..., \sigma^{n}$ are the singular values of 
\begin{equation*}
    \frac{M_a M_b^{T}}{||M_{a}||_{F}||M_{b}||_{F}}~.
\end{equation*}
The Angular Procrustes Distance satisfies the mathematical definitions of a metric, and thus especially the triangle inequality. The Angular Procrustes Similarity is then given by $1 - 2\phi /\pi$. Both similarity measures are used to study representational alignment in artificial neural networks where the data matrices have a shape of (sample, feature). Here, we use these measures for general matrices which motivates our adaptation of CKA. \citet{Cloos24} show that CKA is mainly influenced by the first principal component while the Angular Procrustes Similarity exhibits a higher sensitivity also to other principal components.

\begin{figure*}[!ht]
    \begin{center}
        \includegraphics[height=0.88\textheight]{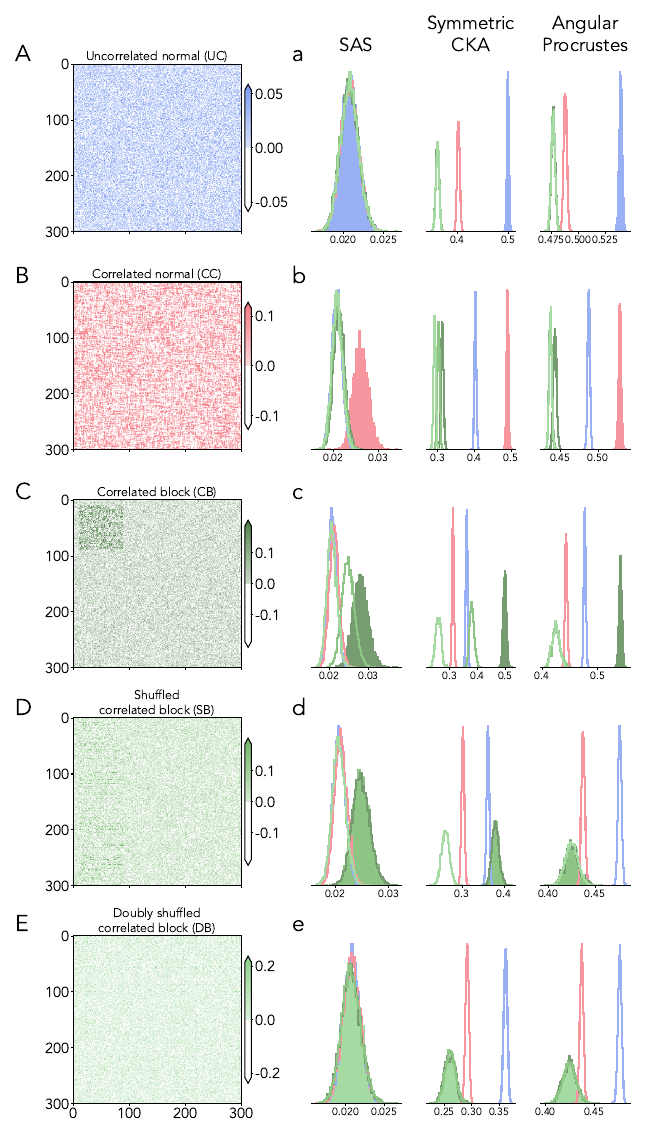}
    \end{center}
    \caption{
    \textbf{Comparing non-square matrices of brain activity with \ac{SAS}, symmetric CKA, and the Angular Procrustes Similarity.} 
    \textbf{A} As in \autoref{fig:comparison}.
    \textbf{B} Histograms of \ac{SAS}, symmetric CKA and the Angular Procrustes Similarity between all individual trials. Filled distributions indicate self-similarities, non-filled ones indicate cross-similarities.}
    \label{supp:comparsion_supp}
\end{figure*}

The comparison between SAS and symmetric CKA as well as the Angular Procrustes Similarity is displayed in \autoref{supp:comparsion_supp}.
For CC and CB matrices, symmetric CKA and the Angular Procrustes Similarity show a larger self- than cross-similarity. This improves the separability of the matrix classes. However, both measures also identify UC matrices as more similar to themselves than to matrices from other classes. This can be considered counterintuitive: by construction, UC matrices should be identified as ``maximally random" in the sense that no excess similarity (compared with other classes) should be identified. Additionally, in all cases, CKA and Angular Procrustes Similarity introduce a hierarchy of similarities assessed by the differences between self- and cross similarities that is difficult to interpret. For example, CKA identifies SB matrices to be more similar to UC matrices than to DB matrices, while the Angular Procrustes Similarity identifies CB matrices to be more similar to UC than to SB matrices. 

In contrast, similarities computed by SAS matches intuitive expectations in the examples studied here. Matrix classes are identified as more similar in the cases where it is plausible (CB and SB). This suggests that SAS is better suited for matrix comparisons that require rigorous statistical testing as maximally random null-models can be constructed and behave as expected.

\subsection{\texorpdfstring{Change in \ac{SAS} under small perturbations}{Change in SAS under small perturbations}}
\label{supp:perturbation}
Consider a matrix $M$ with \ac{SVD}
\begin{equation}
    M = U \Sigma V^{T}~.
\end{equation}
For simplicity, we assume that $M \in \mathbb{R}^{n \times n}$ is a random matrix with non-degenerate singular values.
First, we relate the singular values as well as the left and right singular vectors to the eigenvalues and eigenvectors of certain symmetric matrices. We calculate 
\begin{equation}
    M M^{T} = U \Sigma ^{2} U^{T}, \quad M^{T}M = V \Sigma ^{2} V^{T}
\end{equation}
and observe that $M M^{T}$ and $M^{T}M$ are symmetric, the squared singular values are the eigenvalues of the matrices, and since $U$ and $V$ are orthogonal matrices, the left and right singular vectors are the eigenvectors.
We perturb $M$ by a random $\sqrt{\epsilon} W$, thus considering $M + \sqrt{\epsilon} W$. This induces a perturbation up to first order (in $\sqrt{\epsilon}$) of $\sqrt{\epsilon}\big(MW^{T} + WM^{T}\big)$ in $MM^{T}$ and  $\sqrt{\epsilon}\big(M^{T}W + W^{T}M\big)$ in $M^{T}M$.
Following Rayleigh-Schr{\"o}dinger perturbation theory \citep{Landau_QM} , the perturbed left singular vectors $\tilde{u}^{j}$ read
\begin{equation}
    \tilde{u}^{j} = u^{j} + \sqrt{\epsilon}\sum_{k\neq j} \frac{\langle u^{k}, \big(MW^{T} + WM^{T}\big)u^{j} \rangle}{{\sigma^{j}}^{2}-{\sigma^{k}}^{2}}u^{k} + \epsilon \cdot \textnormal{ terms orthogonal to } u^j~.
\end{equation}
while the perturbed eigenvalues are
\begin{equation}
    \tilde{\sigma}^{{j}^{2}} = \sigma^{{j}^{2}} + \sqrt{\epsilon} \langle u^{j}, \big(MW^{T} + WM^{T}\big)u^{j}\rangle + \epsilon\sum_{k\neq j} \frac{\langle u^{k}, \big(MW^{T} + WM^{T}\big)u^{j} \rangle^2}{{\sigma^{j}}^{2}-{\sigma^{k}}^{2}}~.
\end{equation}

To investigate how the eigenvalues change under perturbation, we calculate:
\begin{align}
    & \langle u^{j}, \big(MW^{T} + WM^{T}\big)u^{j}\rangle \notag \\
    = & \langle u^{j}, MW^{T} u^{j} \rangle + \langle u^{j}, WM^{T}u^{j}\rangle \notag \\
    = &\langle W M^{T} u^{j}, u^{j} \rangle + \langle u^{j}, WM^{T}u^{j}\rangle \notag \\
    = & 2 \langle W M^{T} u^{j}, u^{j} \rangle \notag \\
    = & 2 \sigma^{j} \langle W v^{j}, u^{j} \rangle~.
\end{align}
Similarly, 
\begin{equation}
    \langle u^{k}, \big(MW^{T} + WM^{T}\big)u^{j}\rangle = \sigma^k \langle W v^{k}, u^{j} \rangle + \sigma^j \langle u^{k}, W v^{j} \rangle~.
\end{equation}
Writing the perturbation matrix $W$ as an expansion of the left and right singular vectors, we get
\begin{equation}
    W = \sum_{j,k} \bar{w}_{jk}u^{j}(v^{k})^{T}
\end{equation}
where $\bar{w}_{ij} \in \mathbb{R}$ are the corresponding coefficients. Thus
\begin{equation}
    \tilde{\sigma}^{{j}^{2}} = \sigma^{{j}^{2}} + \sqrt{\epsilon} 2 \sigma^{j} \bar{w}_{jj} + \epsilon\sum_{k\neq j} \frac{\left(\sigma^k \bar{w}_{jk} + \sigma^j \bar{w}_{kj}\right)^{2}}{{\sigma^{j}}^{2}-{\sigma^{k}}^{2}}~.
\end{equation}
Consequently, it is easy to construct perturbations that leave certain singular values unchanged. For a random perturbation, however, this is unlikely: even when the perturbation changes a single entry of $M$, since the left and right singular vectors generally have no resemblance to the standard normal basis of $\mathbb{R}^{n}$, most of the $\bar{w}_{ij}$ differ from zero. Neglecting the special case in which the first and second order exactly cancel, we can assume that all eigenvalues change under perturbation.

Writing
\begin{equation}
    c_{kj} = \frac{\langle u^{k}, \big(MW^{T} + WM^{T}\big)u^{j}\rangle}{{\sigma^{j}}^{2}-{\sigma^{k}}^{2}}~,
\end{equation}
normalizing the eigenvectors yields
\begin{equation}
    \tilde{u}^{j} = \bigg(1 - \frac{\epsilon}{2}\sum_{k\neq j} c_{kj}^{2} \bigg) u^{j} + \sqrt{\epsilon} \cdot \textnormal{ terms orthogonal to } u^j
\end{equation}
and \textit{mutatis mutandis} for the perturbed right singular vectors $\tilde{v}^{j}$. This implies for the angles between the left singular vectors of the unperturbed and perturbed matrix
\begin{align}
    \alpha^{j}(U) & =\arccos \left(\langle u^{j}, \tilde{u}^{j} \rangle \right) \notag\\
                  & = \arccos \left( \langle u^{j}, \bigg(1 - \frac{\epsilon}{2}\sum_{k\neq j} c_{kj}^{2} \bigg) u^{j} + \sqrt{\epsilon} \cdot \textnormal{ terms orthogonal to } u^j \rangle \right)\notag\\
                  & = \arccos \bigg(1 - \frac{\epsilon}{2}\sum_{k\neq j} c_{kj}^{2} \bigg)
\end{align}
where we used the orthonormality of the $u^{i}$.

By a similar argument as for the eigenvalues, we can also assume that for a generic perturbation, the eigenvectors change. Thus, $\sum_{k\neq j }c_{kj}^2 > 0$ since $\langle u^{j}, \tilde{u}^{j} \rangle < 1$, and we denote by $C(M, W, U)$ the smallest of those terms. Hence, we obtain
\begin{equation}
     \alpha^{j}(U) \leq \arccos( 1 - \epsilon C(M, W, U))
\end{equation}
where the constant also absorbs the factor $\frac{1}{2}$.
Since the same relations hold for the right singular vectors, we have $\alpha^{j} \leq \arccos( 1 - \epsilon^{2} C(M, W))$ where $C(M, W)=\min \{C(M, W, U), C(M, W, V)\}$. And as this holds true for all angles $\alpha^{j}$, it implies that \ac{SAS} between the matrices $M$ and $M + \sqrt{\epsilon} W$ is $1 - \arccos(1-\epsilon C(M,W) ) / \frac{\pi}{2}$ up to first order in $\epsilon$.

\subsection{\texorpdfstring{The Perron-Frobenius Theorem and the Perron vector
}{The Perron-Frobenius Theorem and the Perron vector}}
\label{supp:perron}
The Perron-Frobenius theorem \citep{Deuflhard08} for non-negative square matrices asserts that if a matrix $N$ has only non-negative entries, then
\begin{itemize}
    \item there is a real eigenvalue $\lambda_{\mathrm{P}}\geq0$ of $N$ such that, for all other eigenvalues $\lambda$ of $N$, $|\lambda| \leq \lambda_{\mathrm{P}}$, and
    \item the eigenvector corresponding to the eigenvalue $\lambda_{\mathrm{P}}$ has only non-negative entries. This eigenvector is called the Perron vector.
\end{itemize}
For the random graphs in this work, \ac{SAS} considers $N=M_{i}M_{i}^{T}$ and $N=M_{i}^{T}M_{i}$ where $M_{i}$ is an adjacency matrix with non-negative entries. In both cases, $N$ is a symmetric positive-definite matrix. Both $N$ share all non-zero eigenvalues (the squares of the singular values of $M_i$), and the eigenvectors of $N$ are the right and left singular vectors (up to a potential multiplication with $-1$.) Since by assumption non-trivial singular values are unique, and under the assumption that at least one non-trivial singular value exists, the above inequalities become strict, and the eigenvector to $\lambda_{\mathrm{P}}$ becomes unique.

\subsection{Non-square matrices from subsampled networks}
\label{supp:non-square-networks}

We define subsampled graphs as graphs where the connectivity is only known for a random subset of source nodes.
Thus, the adjacency matrices are non-square. We keep the number of possible connections in the network the same by instantiating a full graph of size $N_> = \frac{1}{f} N_N$ where $f < 1$, and then subsample $N_< = f N_N$ source nodes. For all networks we here choose $f = 2/3$ and for the comparison between subsampled graphs in this work, the list of source nodes for which information is available is shared across network instance and models. We use \ac{SAS} to test how similar the structures of subsampled networks are.
\begin{figure*}[!ht]
    \begin{center}
        \includegraphics[width=\textwidth]{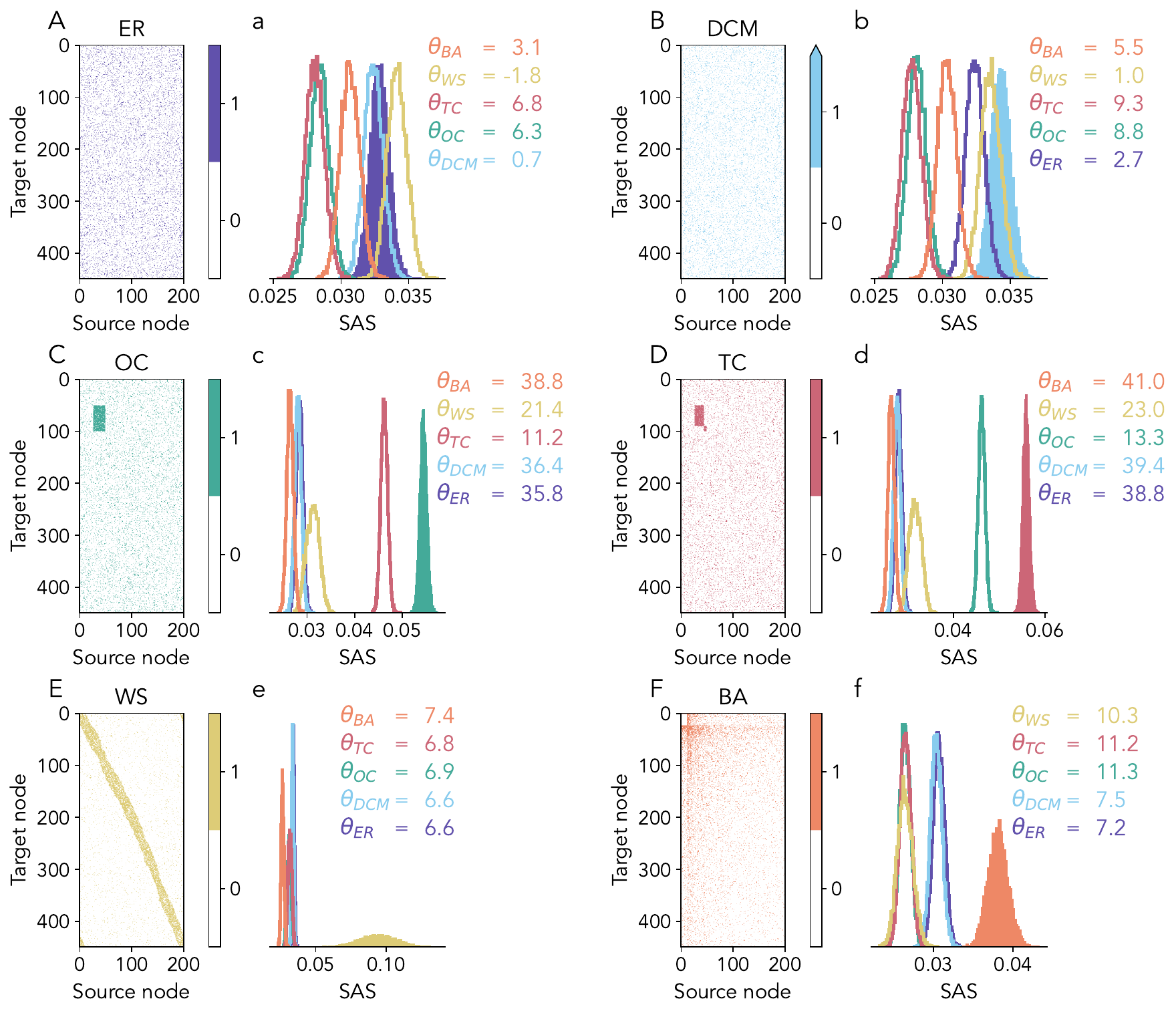}
    \end{center}
    \caption{
    \textbf{Self- and cross-similarity of subsampled network models.}
    \textbf{A--F} Single instances of the different network models. Colored matrix elements indicate a connection between nodes. 
    \textbf{a--f} 
    Histograms of \ac{SAS} between instances of the network models ($n=100$, all pairs compared). Filled distributions indicate self-similarities, non-filled ones indicate cross-similarities. Effect sizes $\theta$ between self- and cross-similarities also shown.}
    \label{supp:connectomes_and_similarity_rectangular}
\end{figure*}
Instances of these subsampled graphs, their respective self- and cross-similarities, and the corresponding effect sizes are shown in \autoref{supp:connectomes_and_similarity_rectangular}. In subsampled graphs, some of the original structure is inevitably invisible, and we therefore expect \ac{SAS} to find less distinguishable structure. This is especially apparent for the \ac{DCM} network, which does not have enough structure to be reliably distinguished from the other network models. All other subsampled network graphs retain enough structure such that \ac{SAS} can distinguish between them.

\subsection{\texorpdfstring{\ac{SAS} for $1\times n$ matrices}{SAS for 1xn matrices}}
\label{supp:1d_case}
Let $v_{i}\in \mathbb{R}^{1\times n}$ where $i \in \{a, b\}$. In this case, the \ac{SVD} reads
\begin{equation}
    v_{i} = 1 \cdot \mathrm{diag}(\lVert v_{i} \rVert, 0, ...,0) \cdot\begin{bmatrix} 
     & \frac{v_{i}}{\lVert v_{i} \rVert}_{2} & \big| & \ast & \big| & \dots & \big| & \ast & \\
     \end{bmatrix} ~.
 \end{equation}
Thus, there is only one non-zero singular value, resulting in the weighted sum used in the definitions of \ac{SAS} to consist of one summand, with weight one. Thus, \ac{SAS} is the angular similarity between $v_{a}$ and $v_{b}$ in this case.

\subsection{Comparison of SAS with common measures on brain data}

\begin{figure*}[!ht]
    \begin{center}
        \includegraphics[width=\textwidth]{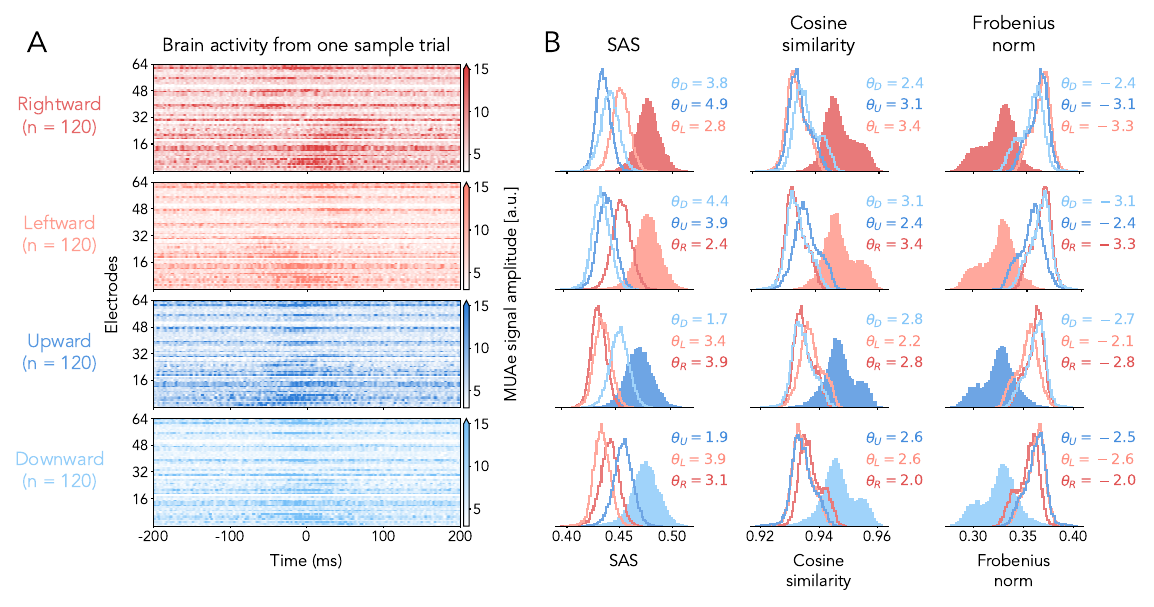}
    \end{center}
    \caption{
    \textbf{Comparing non-square matrices of brain activity with \ac{SAS}, cosine similarity, and the Frobenius norm.} 
    \textbf{A} Brain activity of all recorded electrodes for one example trial of each type (different trials as shown in \autoref{fig:brain_activity}).
    \textbf{B} Histograms of \ac{SAS}, cosine similarity and the Frobenius norm between all individual trials. Filled distributions indicate self-similarities, non-filled ones indicate cross-similarities. Effect sizes $\theta$ for the comparison between self- and cross-similarities also shown.}
    \label{supp:brain_activity}
\end{figure*}

\begin{figure*}[!ht]
    \begin{center}
        \includegraphics[width=\textwidth]{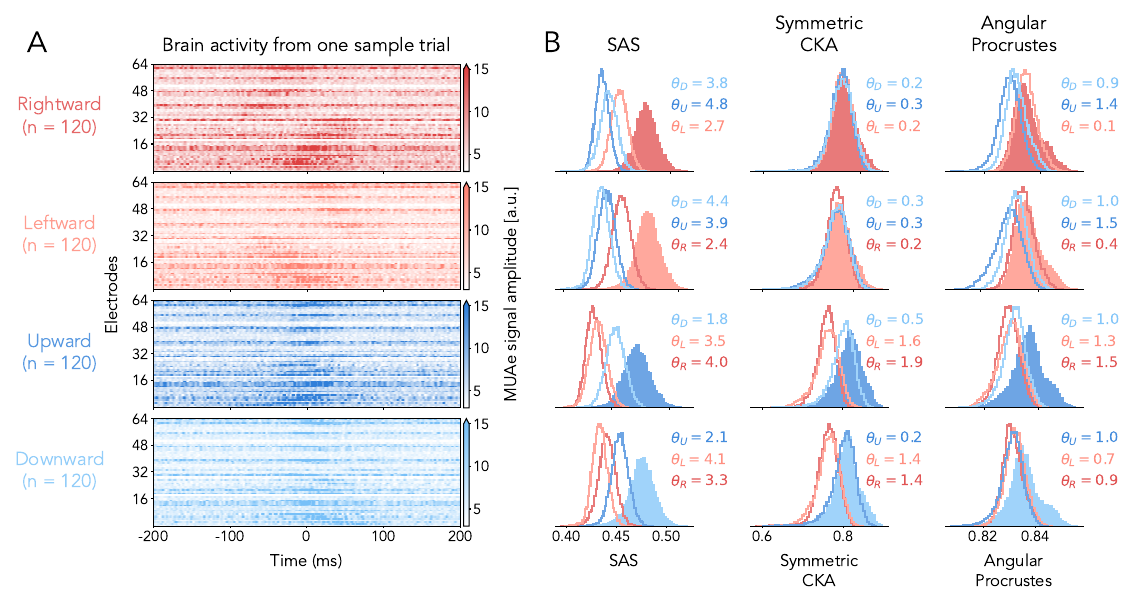}
    \end{center}
    \caption{
    \textbf{Comparing non-square matrices of brain activity with \ac{SAS}, cosine similarity, and the Frobenius norm.} 
    \textbf{A} Brain activity of all recorded electrodes for one example trial of each type (different trials as shown in \autoref{fig:brain_activity}).
    \textbf{B} Histograms of \ac{SAS}, symmetric CKA and the Angular Procrustes Similarity between all individual trials. Rest as in \autoref{supp:brain_activity}.}
    \label{supp:brain_activity_add}
\end{figure*}

We compare how similarity is assessed by \ac{SAS}, cosine similarity, and the Frobenius norm between the different trial types of brain activity data (\autoref{sec:brain_data}). \autoref{supp:brain_activity} shows that only \ac{SAS} identifies that the recorded activity is more similar between trials where the stimulus has the same orientation (U-D and L-R).
Additionally, we use symmetric CKA and the Angular Procrustes Similarity on the brain activity data \autoref{supp:brain_activity_add} (for a definition of the similarity measures see \autoref{supp:rm_supp}). We find that both similarity measures fail to separate the different brain states from one another, except for the Upward state using Angular Procrustes Similarity. This demonstrates the limitation of these measures in the application to time series data.

\end{document}